\documentclass[10pt]{article}
\usepackage[utf8]{inputenc}
\usepackage[english]{babel}
\usepackage{hyperref}
\usepackage{textcomp}
\usepackage[autostyle]{csquotes}

\usepackage[backend=biber,style=ieee, citestyle=numeric-comp,maxbibnames=99]{biblatex}
\AtBeginBibliography{\footnotesize}

\usepackage{amssymb}
\usepackage{amsthm}
\usepackage{amsfonts}
\usepackage{amssymb}
\usepackage{graphicx}
\usepackage{comment}
\usepackage{geometry}
\usepackage{subfiles}
\usepackage{subfig}
\usepackage{algpseudocode}
\usepackage{algorithm}
\usepackage{mathtools}
\usepackage{pgfplots}
\pgfplotsset{/pgf/number format/use comma,compat=newest}
\usepackage{url}
\usepackage{tikz}
\usepackage{notations}
\usepackage{bbm}
\usepackage{floatflt}
\usepackage{subfig}
\usepackage{authblk}

\usetikzlibrary{arrows,decorations.pathmorphing,backgrounds,positioning,fit,petri}

\usepgfplotslibrary{fillbetween}

\newcommand{\E}{\mathbb{E}}
\newcommand{\R}{\mathbb{R}}

\theoremstyle{plain}

\theoremstyle{definition}

\theoremstyle{remark}

\newcommand\iid{\mathrel{\stackrel{\makebox[0pt]{\mbox{\normalfont\tiny iid}}}{\sim}}}

\title{
The Decimation Scheme for Symmetric Matrix Factorization}
\author{Francesco Camilli$^\ast$ and Marc Mézard$^\dagger$}
\affil{\small $^\ast$\emph{The Abdus Salam International Center for Theoretical Physics, Trieste, Italy}\\
$^\dagger$\emph{Bocconi University, Milan, Italy}}

\begin{document}
\maketitle

{
\let\thefootnote\relax\footnotetext{$^\ast$\url{fcamilli@ictp.it}, $^\dagger$\url{marc.mezard@unibocconi.it}}
}

\begin{abstract}
    Matrix factorization is an inference problem that has acquired importance due to its vast range of applications that go from dictionary learning to recommendation systems and machine learning with deep networks. The study of its fundamental statistical limits represents a true challenge, and despite a decade-long history of efforts in the community, there is still no closed formula able to describe its optimal performances in the case where the rank of the matrix scales linearly with its size. In the present paper, we study this extensive rank problem,  extending the alternative 'decimation' procedure that we recently introduced, and carry out a thorough study of its performance. Decimation aims at recovering one column/line of the factors at a time, by mapping the problem into a sequence of neural network models of associative memory at a tunable temperature. Though being sub-optimal, decimation has the advantage of being theoretically analyzable. We extend its scope and analysis to two families of matrices. For a large class of compactly supported priors, we show that the replica symmetric free entropy of the neural network models takes a universal form in the low temperature limit. For sparse Ising prior, we show that the storage capacity of the neural network models diverges as sparsity in the patterns increases, and we introduce a simple algorithm based on a ground state search that implements decimation and performs matrix factorization, with no need of an informative initialization.
\end{abstract}

\tableofcontents

\section{Introduction}
The factorization of a matrix into two, or more, factors represents a building block for many machine learning and inference problems. A well-known instance of it is \emph{dictionary learning} \cite{olshausen1996,OLSHAUSEN1997,DL_Kreutz03,mairal2009}, which aims at representing a matrix as a product of two factor matrices, where the first, called \emph{dictionary}, is very sparse, and the second, called \emph{feature matrix}, has columns that form an over-complete basis of a euclidean space. As a result, each vector stored in the initial matrix is represented as a linear combination of few elements of the feature matrix. Matrix factorization is also at the basis of recommendation systems \cite{PMF_salakhutdinov}, and in general proves to be very effective whenever we want to reconstruct missing elements in a matrix of data, be it an image, a correlation matrix, or a matrix of preferences \cite{Mairal_colorrestoration,Sapiro_review,Image_denoising}. Other applications of matrix factorization include, but are not limited to, sparse principal component analysis \cite{sparse-PCA}, blind source separation \cite{blind_source_sep}, matrix completion \cite{Candes_completion,Candes_Tao_completion}, robust principal component analysis \cite{robust_PCA}

In more specific terms, matrix factorization is the problem of reconstructing the two factors $\bA$, $\bB$ of a matrix $\bA\bB$ from a potentially noisy observation of the latter, say $\bY$. One would like to answer two main questions: \emph{(i)} in what regimes of sizes of $\mathbf{A}$, $\mathbf{B}$ and noise is it possible to reconstruct the two factors (up to a permutation of the lines of $\mathbf{A}$ and the columns of $\mathbf{B}$) ? \emph{(ii)} Do there exist efficient algorithms that achieve a good performance?

In the present paper we focus on symmetric matrix factorization in which the two factors to retrieve are identical. Consider an $N\times P$ matrix $(\xi_{i}^\mu)_{i\leq N}^{\mu\leq P}=\bxi\in\mathbb{R}^{N\times P}$ whose elements are independently and identically distributed according to a given prior probability $P_\xi$, that we suppose to be symmetric, with unit variance and compact support: $\EE\xi=0$, $\EE\xi^2=1$, $|\xi|\leq C$ for some $C>0$. Secondly, let $(Z_{ij})_{i,j\leq N}=(Z_{ji})_{i,j\leq N}=\bZ$ be a Wigner matrix, that is $Z_{ij}=Z_{ji}\iid\mathcal{N}(0,1+\delta_{ij})$. Symmetric matrix factorization can thus be formulated as an inference problem: a Statistician needs to recover $\bxi$ given the noisy observations
\begin{align}\label{eq:channel}
    \bY=\frac{\bxi\bxi^\intercal}{\sqrt{N}}+\sqrt{\Delta}\bZ\,.
\end{align}
The strength of the noise $\bZ$ w.r.t. that of the signal is tuned by $\Delta\geq0$. In the following we will need to single out the $P$ column vectors inside $\bxi$, denoted by $\bxi^\mu$, and we shall refer to them as \emph{patterns}. Despite the model is presented here in a stylized way, i.e. with the two factors being identical and with completely factorized prior, we believe this setting represents a fundamental first step in the understanding of the general problem. Concerning in particular the assumption of a factorized prior, this is often used also in concrete situations. Indeed, for instance, the $L^2$ norm regulators appearing in the empirical risk used to train neural networks are inherited from a zero temperature limit of a Statistical Mechanics problem that has the empirical risk as a Hamiltonian with factorized prior on the weights of the network, as clarified by \cite{repr_lerning_review}.

A very popular setting to tackle an inference problem is the Bayes-optimal one, in which the Statistician tasked with the reconstruction of $\bxi$ knows the generating process of the observations $\bY$, namely they know that $\bZ$ is Gaussian, they know $N,P,\Delta$ and the probability distribution of factors $P_\xi$. This Bayes-optimal setting is of utmost relevance as it provides the information-theoretic optimal performance. Indeed, the posterior mean estimator $\E[\bX\bX^\intercal|\bY]$, where
\begin{align}
    \label{eq:posterior_dictionaryfull_intro}
    dP(\bxi=\bX\mid\bY)=\frac{1}{\mathcal{Z}(\bY)}\prod_{i\leq N,\mu\leq P}dP_\xi(X_i^\mu)\exp\Big[\frac{1}{2\sqrt{N}\Delta}\Tr\bY\bX\bX^\intercal-\frac{1}{4\Delta N}\Tr(\bX\bX^\intercal)^2\Big]\, ,
\end{align}
is the one that minimizes the mean square error loss on the reconstruction of $\bxi\bxi^\intercal$. The normalization of the distribution $\mathcal{Z}(\bY)$ is called \emph{partition function} and the associated \emph{free entropy} is defined as
\begin{align}
    \label{eq:free_entropy_full}
    \Phi_{N,P}=\frac{1}{NP}\E\log\mathcal{Z}(\bY)\,.
\end{align}
The free entropy has a central role. In fact, from the thermodynamic point of view, it can be used to identify what macrostates dominate probability and are thus selected at thermodynamic equilibrium. These macrostates are usually identified by the values of some global order parameters, such as $\Tr \bX\bX^\intercal\bxi\bxi^\intercal/N^2$, which measures the average alignment of a sample from the posterior and the ground truth $\bxi$ we want to estimate. On the other hand, the free entropy is in close relationship with the \emph{mutual information} $I(\bxi;\bY)$ between the data and the ground truth. This information theoretic quantity quantifies the amount of residual information about the ground truth that is still available in the data after they have been corrupted by the noise.

If the rank $P$ is finite, the model \eqref{eq:channel} is typically referred to as \emph{spiked Wigner model}, first introduced as model for Principal Component Analysis (PCA) \cite{Johnstone_WSM}. The spectral properties of low rank perturbations of high-rank matrices (such as the Wigner matrix $\bZ$) are by now largely understood in random matrix theory, and they can give rise to the celebrated BBP  carry out a thorough study of carry out a thorough study of transition \cite{BBP}, further studied and extended in \cite{baik2006eigenvalues,rmt-Peche2006,feral2007largest,capitaine2009largest,Nadakuditi_Jacobi,BENAYCHGEORGES2011,benaych2012singular,bai2012sample}. Thanks to the effort of a wide interdisciplinary community, we also have a control on the asymptotic behaviour of the posterior measure \eqref{eq:posterior_dictionaryfull_intro} and an exact formula for the free entropy associated to the low-rank problem \cite{lesieur2015mmse,Lelarge2017FundamentalLO,wigner-wishart,Barbier_2019,adaptive,Jean-lenka2,ElAlaoui,camilli2023central} (recently extended to rotational invariant noise \cite{myPNAS}), which yields the Bayes-optimal limit of the noise allowing the reconstruction of the low-rank spike. Finally, a particular class of algorithms, known as \emph{Approximate Message Passing} (AMP) \cite{Mezard_1989,KabashimaAMP,donohoAMP,Rangan_iterative18,AMP_statevolution}, is able to perform factorization up to this Bayes-optimal limit.

Here we are interested in the extensive rank regime where $P,N\to\infty$ with fixed ratio $P/N=\alpha$. In the hypothesis of a rotationally invariant noise $\bZ$, the spectral properties of $\bY$ are governed by the free-convolution \cite{VOICULESCU1986323} of the spectral densities of $\bZ$ and $\bxi\bxi^\intercal$. On the information theoretic side instead, there still is no accepted closed formula that expresses $\Phi_{N,P}$. Hence, the information theoretic limits are currently out of reach, and the Minimum Mean Square Error (MMSE) for this estimation problem is not known. Among the past attempts, we must mention the line of works \cite{Marc-Kabashima,Parker-AMP1,Parker-AMP2,ZZY21,Lucibello_2022}, whose proposed solution, as pointed out in \cite{schmidt:tel-03227132,perturbative_Maillard21}, provides only an approximation of the correct limit. In fact, the authors of \cite{perturbative_Maillard21} build a perturbative approach that highlights the presence of relevant correlations neglected in the previous works. A further attempt to produce a closed replica formula was put forward in \cite{barbier2022DL}, but, as \cite{Marc-Kabashima}, it involves uncontrolled approximations. 

The main obstacle in the computation of the asymptotics of \eqref{eq:free_entropy_full} is the fact that it is a matrix model, and, in particular, the term $\Tr(\bX\bX^\intercal)^2$ couples both the \enquote{rank, or patterns indices} $\mu$, and the \enquote{dimension, or particle site indices} $i$. We will use here a different approach that we introduced and studied recently \cite{MFNN} in the simplest case where the factors' elements $\xi_i^\mu$ are independent binary variables. Instead of the Bayes-optimal setting we use a simpler procedure, that we call \emph{decimation}. At the cost of giving up on Bayes-optimality, decimation solves this problem and allows us to identify an iterative scheme to estimate pattern by pattern, giving an estimate of $\bxi$ through a sequential estimation of its columns, and, more importantly, whose asymptotic performance turns out to be completely analyzable. In the case of binary patterns we could thus show that matrix factorization is possible in a part of the phase diagram where $\alpha$ and $\Delta$ are small enough. Here we generalize this approach to arbitrary distributions of the patterns' elements.

\paragraph{Organization of the paper and main contributions} In Section \ref{sec:Decimation} we define the decimation scheme, laying the ground for the replica computation of Section \ref{sec:free_entropy}. In Section \ref{0T_section}, we compute the low temperature limits for two classes of priors: sparse Ising and a generic absolutely continuous, symmetric and bounded support prior. Surprisingly, the free entropies of the neural network models arising from decimation evaluated at the equilibrium value of the order parameters have a universal form, but in general not the same numerical value. 

As we shall argue in the following, the starting point of the decimation procedure, i.e. the initial value of the parameters $\alpha$ and $\Delta$, is of crucial importance for its success. Therefore, in Section \ref{sec:diagrams} we analyze the phase diagrams for the initial step of decimation. For the sparse Ising prior, we show that as sparsity increases, the storage capacity of the sequential neural network models of decimation diverges. For the class of continuous priors we highlight the presence of a thermodynamic transition, where there is a non-trivial overlap between a sample from the Gibbs measure and the sought pattern, and a performance transition, where Gibbs sampling can outperform the null-estimator.

In Section \ref{sec:numerical_tests} we provide numerical evidence in support of the replica theory. We introduce the Decimated AMP algorithm (DAMP), in order to verify the predictions of the replica theory, and we relate the replica symmetric order parameters to the mean square error on the reconstruction of the patterns, as well as to the matrix mean square error for matrix denoising, showing that decimation can outperform Rotational Invariant Estimators (RIEs) \cite{RIE_bouchaud,Troiani2022OptimalDO,pourkamali2023rectangular} in this task. Furthermore, this Section contains the pseudo-code of a ground state oracle, an algorithm that is indeed able to find all the patterns one by one, with no need of informative initialization, contrary to DAMP. 

Section \ref{sec:related_works} contains a comparison with recent relevant works that are related to the present one. Finally, Section \ref{sec:conslusions} gathers the conclusions and future perspectives.

\section{Decimation}\label{sec:Decimation}
Let us give a closer look at the probability distribution \eqref{eq:posterior_dictionaryfull_intro}. For the purpose of the theoretical analysis we can replace $Y_{ij}$ with the r.h.s. of \eqref{eq:channel}, getting
\begin{align}
    \label{eq:posterior_dictionaryfull}
    dP(\bxi=\bX\mid\bY)&=\frac{1}{\mathcal{Z}(\bY)}\prod_{i\leq N,\mu\leq P}\left[dP_\xi(X_i^\mu)\right]
    \e^{-\beta \left[\sum_\mu (E_1(\bX^\mu)+E_2(\bX^\mu)+E_3(\bX^\mu))+\sum_{\mu <\nu} E_4(\bX^\mu,\bX^\nu))\right]}
\end{align}
where $\beta=\frac{1}{\Delta}$, $\bX^\mu=(X^\mu_i)_{i\leq N}$ and
\begin{align}
E_1(\bx)&=-\sum_{i,j=1}^N J_{ij} x_i x_j\ \ ; \ \ J_{ij}=\frac{1}{N}\sum_\nu \xi_i^\nu \xi_j^\nu\\
E_2(\bx)&=-\sum_{i,j=1}^N \frac{\sqrt{\Delta}}{2\sqrt{N}} Z_{ij}x_i x_j\\
E_3(\bx)&=\frac{1}{4N}\Big[\sum_i x_i^2\Big]^2\\
E_4(\bx,\bx')&= \frac{1}{2N}\Big[\sum_i x_i x'_i\Big]^2\,.
\label{4Edef}
\end{align}
Here one should be careful not to confuse $\xi_i^\mu$ which is the 'ground-truth' matrix from which the signal $\bY$ was generated, and $X_i^\mu$ which is a random variable distributed according to the measure $dP(\bxi=\bX\mid\bY)$, so that the expectation value of $X_i^\mu$ gives the best possible approximation to $\xi_i^\mu$.

Looking at the above decomposition, we notice that, if we could drop the term $E_4(\bX^\mu,\bX^\nu)$, we would have a system of $P$ decoupled problems, one for each value of $\mu$, described by an energy $E_1(\bX^\mu)+E_2(\bX^\mu)+E_3(\bX^\mu)$. The energy $E_1$ is that of a spin glass with $N$ variables $x_i$, each with an a-priori measure $P_\xi(x_i)$, interacting by pairs through a matrix of couplings $J_{ij}$ which has a Hebbian form determined by the ground-truth patterns $\bxi$. The energy $E_2$ is a random spin glass term created by measurement noise. The energy $E_3$ is a global penalty that ensures that the norm of $\bX$ does not get too large; one can also incorporate it into the local measure using a Lagrange multiplier. Altogether, the system described by $E_1+E_2+E_3$ is a spin glass Hamiltonian with an interaction which is a noisy version of a Hebbian interaction. This is typical of problems that have been studied as neural networks for associative memory, following the seminal work by Hopfield \cite{Hopfield82}. The present one is a generalization of the Hopfield model, where the stored patterns components $\xi_i^\mu$ are no longer binary but have a more general distribution which can be continuous. Based on our knowledge of associative memories, one can expect that, when the noise strength $\Delta$ and the number of patterns per variable $\alpha=P/N$ are small enough, there can exist a 'retrieval' phase, in which the configurations $\bx$ that minimize $E_1(\bx)+E_2(\bx)+E_3(\bx)$ are close to the stored patterns $\xi_i^\mu$. This is certainly the case for binary patterns as shown in \cite{MFNN}. Assuming that such a retrieval phase exists, one can understand the use of the fourth energy term, $E_4$. In fact one can interpret \eqref{eq:posterior_dictionaryfull_intro} as follows: we start from $P$ replicas of an associative memory each with energy $E_1(\bX^\mu)+E_2(\bX^\mu)+E_3(\bX^\mu)$. These copies interact by pairs through the term $E_4(\bX^\mu,\bX^\nu)$ which is a repulsive term. If one works in the retrieval phase of the associative memory, then at low temperature the ground state will be found when each replica $\bX^\mu$ is close to one of the patterns $\bxi^{\pi(\mu)}$. As there are $P$ retrieval states and $P$ replicas, all the $\pi(\mu)$ must be distinct from one another, and therefore $\pi$ is a permutation. In such a scenario, one would have found a phase where the factors can be reconstructed.

Decimation is based precisely on this idea. It works as a sequence of $P$ estimations, each one studying a probability distribution which is that of a neural network model of associative memory. More precisely, one looks for one column $\bxi^\mu$ of $\bxi$ at a time. 

To fix ideas, let us start by discussing the search of a first pattern, using a Gibbs measure in the form
\begin{align}
    dP(\bx\mid\bY)=\frac{dP_\xi(\bx )}{\mathcal{Z}_0(\bY)} \exp\Big(\beta\Big[\frac{1}{2N}\sum_{\mu=1}^P\Big(\sum_{i=1}^N\xi_i^\mu x_i\Big)^2+
    \frac{\sqrt{\Delta}}{2\sqrt{N}}\sum_{i,j=1}^NZ_{ij}x_ix_j
    -\frac{\Vert\bx\Vert^4}{4N}
    \Big]\Big)\,.
\end{align}Here we have introduced a factor $\beta$ that plays the role of an inverse absolute temperature for this Boltzmann-Gibbs measure. We could use $\beta=1/\Delta$ as in the Bayes-optimal approach, but as we shall see taking the large $\beta$ limit can also be a good choice.

When using this approach with variables $x_i$ that are not constrained on the hypercube $\{-1,1\}^N$ or in general on a sphere, it is also useful to introduce another term in the exponential that favours $\bx$-configurations with square norm equal to $N$, as we know that the original signal is centered and with unit variance. Hence, the Boltzmann-Gibbs measure that we use to find a first pattern is actually $ dP_\xi(\bx) e^{-\beta E(\bx|\bY)}/\mathcal{Z}_0$ with an energy function
\begin{align}\label{eq:energy_firststep}
    -E(\bx|\bY)=\frac{\sqrt{\Delta}}{2\sqrt{N}}\sum_{i,j=1}^NZ_{ij}x_ix_j+\frac{N}{2}\sum_{\mu=1}^P (m^\mu(\bx))^2
    -\frac{\Vert\bx\Vert^4}{4N}-\frac{\lambda}{4N}(\Vert \bx\Vert^2-N)^2
\end{align}
where we have introduced the \emph{Mattis magnetization}
\begin{align}
    \label{eq:mattis_mag_def}
    m^\mu(\bx)=\frac{1}{N}\sum_{i=1}^N\xi_i^\mu x_i\,.
\end{align}
$\lambda$ is a parameter penalizing (if positive) configurations with $\Vert\bx\Vert^2\neq N$, as mentioned before. If $\lambda\to+\infty$ then the spins are constrained on a sphere. Let us now assume that we are able to sample a configuration $\boldsymbol{\eta}^P$ from the Boltzmann-Gibbs measure with energy \eqref{eq:energy_firststep} that, without loss of generality (we shall relabel the patterns in such a way  that the permutation $\pi$ is the identity), we take as an estimate of $\bxi^P$. How do we find the estimate of the other $\bxi^\mu$, $\mu<P$?

If $\boldsymbol{\eta}^P$ is a good estimate of $\bxi^P$, the corresponding rank one contribution $\boldsymbol{\eta}^P\boldsymbol{\eta}^{P\intercal}$ should be close (in Frobenius norm) to $\bxi^P\bxi^{P\intercal}$. Then, if we subtract it from the Hebbian coupling $E_1(X)$, we can hope that the ground state of the new associative memory problem will now have only $P-1$ ground states, each close to one of the patterns $\bxi^\mu$, $\mu=1,...,P-1$. This new associative memory problem therefore has $P-1$ stored patterns instead of $P$ so that the well known phenomenon of \emph{pattern interference} \cite{Amit0,Amit1}, which limits the storage capacity, will be reduced. 

Based on this intuition, we define the decimation procedure as follows: after having found the first estimate of a pattern, we modify the coupling matrix as 
\begin{align}
    \bY_1=\bY-\frac{\boldsymbol{\eta}^P{\boldsymbol{\eta}^{P\intercal}}}{\sqrt{N}}\,,
\end{align}
which gives a modified energy function
\begin{align}
    \label{eq:energy_secondstep}
    -E(\bx|\bY_1)=\frac{\sqrt{\Delta}}{2\sqrt{N}}\sum_{i,j=1}^NZ_{ij}x_ix_j+\frac{N}{2}\sum_{\mu=1}^P (m^\mu(\bx))^2-\frac{N}{2}(p^P(\bx))^2
    -\frac{\Vert\bx\Vert^4}{4N}-\frac{\lambda}{4N}(\Vert \bx\Vert^2-N)^2
\end{align}
where, here and in the following
\begin{align}
    p^\mu(\bx)=\frac{1}{N}\sum_{i=1}^N\eta_i^\mu x_i\,.
\end{align}
The same reasoning as above applies to this second step.

In general, if the first $R$ ($=0,1,2,\dots,P-1$) patterns have already been estimated, the decimation assumes to produce the estimate of the $R+1$-th pattern sampling from the Boltzmann Gibbs measure
\begin{align}\label{eq:generic_step_BGmeasure}
    d\mu_R(\bx)=\frac{dP_\xi(\bx)}{\mathcal{Z}_R}\exp\big(-\beta E(\bx|\bY_R)\big)
\end{align}where
\begin{align}
    \label{eq:modified_obs}
    \bY_R=\bY-\sum_{\mu=P-R+1}^P\frac{\boldsymbol{\eta}^\mu\boldsymbol{\eta}^{\mu\intercal}}{\sqrt{N}}
\end{align}and
\begin{align}
    \label{eq:generic_step_energy}
    -E(\bx|\bY_R)=\frac{\sqrt{\Delta}}{2\sqrt{N}}\sum_{i,j=1}^NZ_{ij}x_ix_j+\frac{N}{2}\sum_{\mu=1}^P (m^\mu(\bx))^2-\frac{N}{2}\sum_{\mu=P-R+1}^P(p^\mu(\bx))^2
    -\frac{\Vert\bx\Vert^4}{4N}-\frac{\lambda}{4N}(\Vert \bx\Vert^2-N)^2\,.
\end{align}
The energy function above has some desirable features. First, the summation of the squared Mattis' magnetizations attracts mass of the distribution towards those configurations that are most aligned with one of the columns of $\bxi$, which are our goal. Secondly, if the $R$ estimates $\boldsymbol{\eta}^\mu$, with $\mu=P-R+1,\dots P$ are reliable, in a sense we shall specify later, the summation containing the squared $(p^\mu(\bx))^2$ repels the mass of the probability distribution from those configurations that are similar to previously estimated patterns, preventing the sampling from finding a pattern more than once.  

We notice at this point that there are three noise sources in this procedure:
\begin{itemize}
    \item [(a)] the original Wigner matrix $\bZ$;
    \item [(b)] pattern interference whose strength, as discussed above, is increasing with the ratio $\alpha=P/N$;
    \item [(c)] the imperfect retrieval of patterns in the previous steps of decimation.
\end{itemize}
(c) is maybe the least obvious one. At each step, we  subtract a rank one contribution $\boldsymbol{\eta}^\mu\boldsymbol{\eta}^{\mu\intercal}/\sqrt{N}$ that is not exactly $\bxi^\mu\bxi^{\mu\intercal}/\sqrt{N}$. This introduces an additional form of noise that  depends on the quality of the previous reconstructions.

In order to monitor the strength of this third noise, we introduce the \emph{retrieval accuracy} of a pattern $\bxi^\mu$:
\begin{align}
    \label{eq:retrieval_acc}
    m^\mu=\frac{\bxi^\mu\cdot\boldsymbol{\eta}^\mu}{N}\,,\quad \mu=P-R+1,\dots,P\,.
\end{align}
These quantities turn out to be order parameters of the previous decimation steps. Indeed, they are nothing but Mattis' magnetizations of typical samples from \eqref{eq:generic_step_BGmeasure} with a pattern. Hence, each decimation step has its own free entropy and we will determine the new retrieval accuracy via consistency equations arising from the maximization of it, namely we look for those macrostates that dominate probability in the $N\to\infty$ limit. In addition to $m^\mu$ we will have other order parameters appearing. In particular, there will be one, denoted by $r$, tuning the amplitude of the overall noise, that, according to the considerations above, must comprise the three contributions coming from sources (a), (b) and (c). 

\subsection{An assumption on retrieval accuracy}
In order to carry out the computations we need some information on the statistics of the retrieved configurations $\boldsymbol{\eta}^\mu$. We assume that an \enquote{oracle} algorithm will produce $\boldsymbol{\eta}^\mu$ with an asymptotic measure given by
\begin{align}
    \label{local_measure}
    \eta_i^\mu\,\sim\,\langle\cdot\rangle_{\xi_i^\mu,Z}=\frac{\int dP_\xi(x)e^{(Z\sqrt{r}+\beta m^\mu\xi_i^\mu)x
    -\frac{r+u}{2}x^2}(\cdot)}{\int dP_\xi(x)e^{(Z\sqrt{r}+\beta m^\mu\xi_i^\mu)x
    -\frac{r+u}{2}x^2}}\,,\quad \xi_i^\mu\sim P_\xi\,, Z\sim\mathcal{N}(0,1)\text{ independent of other noises}\,,
\end{align}where $m^\mu$, \emph{i.e.} the retrieval accuracy for $\boldsymbol{\eta}^\mu$, and $\,r,\,u$ must be determined self-consistently. \eqref{local_measure} amounts to requiring that, asymptotically, the sites are decoupled and they feel an effective external random magnetic field, that is Gaussian with a mean shifted by the ground truth $\xi_i^\mu$.
Define for later convenience the quantities
\begin{align}
    &\mathbb{E}_{\boldsymbol{\eta}|\boldsymbol{\xi}} [\eta_i^\mu]=m_i^\mu\,,\quad \mathbb{E}_{\boldsymbol{\eta}|\boldsymbol{\xi}}[(\eta_i^\mu)^2]=v_i^\mu\,.
\end{align}Then \eqref{local_measure} has the following implications:
\begin{align}\label{oracle_properties_general}
    &\mathbb{E}_{\boldsymbol{\xi}}[\eta_i^\mu]=\mathbb{E}_{\boldsymbol{\xi}}\mathbb{E}_{\boldsymbol{\eta}|\boldsymbol{\xi}} [\eta_i^\mu]=0\,,\quad \mathbb{E}_{\boldsymbol{\xi}}[\xi_i^\mu m_i^\nu]=m^\mu\delta_{\mu,\nu}\,,\quad \mathbb{E}_{\boldsymbol{\xi}}[v_i^\mu]=v^\mu
\end{align}
that will be self-consistent with the fixed point equations for each decimation step. We shall see from the replica computation that this assumption holds inductively: if it is true at the $R$-th decimation step, then we are able to decouple the site indices also for the step $R+1$, and the resulting spin-glass model has an effective random magnetic field of the same form.

\section{Decimation free entropies}\label{sec:free_entropy}
In this section we compute the large $N$ limit of the free entropy
\begin{align}\label{free_entropy_def}
    \Phi=\lim_{N\to\infty} \frac{1}{N}\mathbb{E}\log\int dP_\xi(\mathbf{x})\exp\left[-\beta E(\mathbf{x}|\mathbf{Y}_R)\right]\,,
\end{align}
where $\mathbb{E}$ is taken w.r.t. all the disorder: $\mathbf{Z},\boldsymbol{\xi},\boldsymbol{\eta}$, and recall that $R$ is the number of patterns that were already estimated. This is done using the \emph{replica method} \cite{MPV}.
We thus introduce
\begin{align}
    \label{replicated_annealed_partfunc}
    \mathbb{E}\mathcal{Z}^n_N:=\mathbb{E}_{\mathbf{Z}}\mathbb{E}_{\boldsymbol{\xi},\boldsymbol{\eta}}\int\prod_{a=1}^n dP_\xi(\mathbf{x}_a)\exp\left[-\beta\sum_{a=1}^nE(\mathbf{x}_a|\mathbf{Y_R})\right]\,.
\end{align}
We decompose this computation and start with the first noise terms in \eqref{eq:generic_step_energy}, and the related $\mathbb{E}_\mathbf{Z}$ average
\begin{multline}
        \mathbb{E}_\mathbf{Z}\exp\left(\frac{\beta\sqrt{\Delta}}{2\sqrt{N}}\sum_{i,j=1}^N Z_{ij}\sum_{a=1}^n x_{a,i}x_{a,j}\right)=
        \exp\left(\frac{\beta^2\Delta}{4N}\sum_{i,j=1}^N \sum_{a,b=1}^n x_{a,i}x_{a,j}x_{b,i}x_{b,j}\right)=\\
        =\exp\left(\frac{N\beta^2\Delta}{4}\sum_{a\neq b}^nQ^2(\mathbf{x}_a,\mathbf{x}_b)+\beta^2\Delta\frac{\Vert\mathbf{x}_a\Vert^4}{4 N}\right)\,.
\end{multline}
where $Q(\bx,\bx')=(1/N)\sum_i x_i x_i'$. For future convenience, we introduce the \enquote{decimation time} $t=R/P$, i.e. the fraction of patterns already estimated. Now we take care of the penalizing $p$-terms in \eqref{eq:generic_step_energy}. After replicating, their contribution to the partition function is
\begin{align}
    \begin{split}
        A:=\prod_{\mu=P(1-t)+1}^P\prod_{a=1}^n e^{-\frac{N\beta}{2}(p^\mu(\mathbf{x}_a))^2}=\prod_{\mu=P(1-t)+1}^P\prod_{a=1}^n\int\frac{ds_a^\mu}{\sqrt{2\pi}}e^{-\frac{(s_a^\mu)^2}{2}+i\sqrt{\frac{\beta}{N}}s_a^\mu\sum_{j=1}^N\eta_j^\mu x_{a,j}}\,.
    \end{split}
\end{align}
Notice that, thanks to the introduction of the auxiliary Gaussian variables $(s_{a}^\mu)_{a\leq n,P(1-t)< \mu\leq P}$, the exponential is now decoupled over the particle indices $j$. Consider then the expectation of $A$ w.r.t. $\boldsymbol{\eta}$, given $\boldsymbol{\xi}$ with the assumptions \eqref{oracle_properties_general}:
\begin{align}
    \begin{split}
        \mathbb{E}_{\boldsymbol{\eta}|\boldsymbol{\xi}}[A]
        =\prod_{\mu=P(1-t)+1}^P\prod_{a=1}^n\int\frac{ds_a^\mu}{\sqrt{2\pi}}\exp\left(-\frac{(s_a^\mu)^2}{2}+\sum_{i=1}^N\log\mathbb{E}_{\eta^\mu_i|\xi^\mu_i}e^{i\sqrt{\frac{\beta}{N}}\eta_i^\mu\sum_{a=1}^n s_a^\mu x_{a,i}}\right)\,.
    \end{split}
\end{align}
Now we can expand the exponential inside the $\log$ up to second order, the remaining terms will be of sub-leading order and thus neglected in the following:
\begin{multline}
    \mathbb{E}_{\boldsymbol{\eta}|\boldsymbol{\xi}}[A]=\prod_{\mu=P(1-t)+1}^P\prod_{a=1}^n\int\frac{ds_a^\mu}{\sqrt{2\pi}}\exp\left(-\frac{(s_a^\mu)^2}{2}+\sum_{a=1}^nis_a^\mu\sqrt{\frac{\beta}{N}}\sum_{i=1}^Nm_i^\mu x_{a,i}-\frac{\beta}{2}\sum_{a,b=1}^ns_a^\mu s_b^\mu\sum_{i=1}^N\frac{(v_i^\mu-(m_i^\mu)^2)}{N}x_{a,i}x_{b,i}\right)\\
    =\prod_{\mu=P(1-t)+1}^P\prod_{a=1}^n\int\frac{ds_a^\mu}{\sqrt{2\pi}}\exp\left[-\frac{1}{2}\sum_{a,b=1}^ns_a^\mu s_b^\mu\left(\delta_{ab}+\beta\sum_{i=1}^N\frac{(v_i^\mu-(m_i^\mu)^2)}{N}x_{a,i}x_{b,i}\right)+\sum_{a=1}^n is_a^\mu\sqrt{\frac{\beta}{N}}\sum_{i=1}^Nm_i^\mu x_{a,i}\right]\,.
\end{multline}
To continue, we  assume condensation on a finite number of patterns, say the first $k$. We focus now on the remaining ones, namely for $\mu>k$:
\begin{align}
    B:=\exp\left[\frac{\beta N}{2}\sum_{a=1}^n\sum_{\mu=k+1}^P(m^\mu(\mathbf{x}_a))^2\right]
    =\int\prod_{\mu=k+1}^P\prod_{a=1}^n\frac{dz_a^\mu}{\sqrt{2\pi}}\exp\left[-\sum_{a=1}^n\sum_{\mu=k+1}^P\frac{(z_a^\mu)^2}{2}+\sqrt{\frac{\beta}{N}}\sum_{a=1}^n\sum_{\mu=k+1}^P z_a^\mu\sum_{i=1}^N x_{a,i}\xi_i^\mu\right]\,.
\end{align}
Putting $A$ and $B$ together, their overall average over $(\boldsymbol{\xi}^{\mu})_{\mu>k}$ takes the form
\begin{multline}
    \mathbb{E}_{(\boldsymbol{\xi}^{\mu})_{\mu>k}}[AB]=\int\prod_{\mu=P(1-t)+1}^P\prod_{a=1}^n\frac{ds_a^\mu}{\sqrt{2\pi}}\int\prod_{\mu=k+1}^P\prod_{a=1}^n\frac{dz_a^\mu}{\sqrt{2\pi}}e^{-\sum_{a=1}^n\left(\sum_{\mu=P(1-t)+1}^P\frac{(s_a^\mu)^2}{2}+\sum_{\mu=k+1}^P\frac{(z_a^\mu)^2}{2}\right)}\\
    \exp\left[
    \sum_{i=1}^N\sum_{\mu=k+1}^P\log\mathbb{E}_{\xi_i^\mu}e^{\sqrt{\frac{\beta}{N}} \sum_{a=1}^nx_{a,i}(\xi_i^\mu z_a^\mu+i\theta(\mu-P+R)m_i^\mu s_a^\mu)-\theta(\mu-P+R)\sum_{a,b=1}^n s_a^\mu s_b^\mu \frac{\beta(v_i^\mu-(m_i^\mu)^2) x_{a,i} x_{b,i}}{2 N}}
    \right]\,,
\end{multline}
where $\theta$ is Heaviside's step function.
If we call $\mathbb{E}_{\boldsymbol{\xi}}m_{i}^{\mu\,2}=:\bar{M}^{\mu\,2}$, a further expansion of the exponential yields:
\begin{align}
    \begin{split}
        &\mathbb{E}_{(\boldsymbol{\xi}^{\mu})_{\mu>k}}[AB]=\int\prod_{\mu=P(1-t)+1}^P\prod_{a=1}^n\frac{ds_a^\rho}{\sqrt{2\pi}}\exp\left[-\frac{1}{2}\sum_{\mu=P(1-t)+1}^P\mathbf{s}^\mu\cdot\left(\mathbbm{1}+\beta(v_{\tau^\mu}-\bar{M}^{\mu\,2})Q\right)\mathbf{s}^\mu\right]\\
        &\int\prod_{\mu=k+1}^P\prod_{a=1}^{n}\frac{dz_a^\mu}{\sqrt{2\pi}}\exp\left\{-\sum_{\mu=k+1}^P\sum_{a=1}^n\frac{(z_a^\mu)^2}{2}+
        \frac{\beta}{2}\sum_{\mu=k+1}^P\sum_{a,b=1}^{n}
        z_a^\mu z_b^\mu Q(\mathbf{x}_a,\mathbf{x}_b)+\right.\\
        &\left.+i\beta\sum_{\mu=P(1-t)+1}^P\mathbb{E}_{\boldsymbol{\xi}}[\xi_1^\mu m_1^\mu]\sum_{a,b=1}^n z_a^\mu s_b^\mu Q(\mathbf{x}_a,\mathbf{x}_b)-\frac{\beta}{\Delta}\sum_{\mu=P(1-t)+1}^P\sum_{a,b=1}^n(\bar{M}^\mu)^2 s_a^\mu s_b^\mu Q(\mathbf{x}_a,\mathbf{x}_b)
        \right\}
    \end{split}
\end{align}
We can now perform a Gaussian integration over the variables $\mathbf{z}^\mu=(z_a^\mu)_{a\leq n}$:
\begin{align}
    \begin{split}
        \mathbb{E}_{(\boldsymbol{\xi}^{\mu})_{\mu>k}}[AB]&=\int\prod_{\mu=P(1-t)+1}^P\prod_{a=1}^n\frac{ds_a^\rho}{\sqrt{2\pi}}\exp\left[-\frac{1}{2}\sum_{\mu=P(1-t)+1}^P\mathbf{s}^\mu\cdot\left(\mathbbm{1}+\beta v^\mu Q+\beta^2 Q
        \frac{\mathbb{E}^2_{\boldsymbol{\xi}}[\xi_1^\mu m_1^\mu]}{\mathbbm{1}-\beta Q}
        Q\right)\mathbf{s}^\mu\right]\\
        &\times\exp\left[-\frac{\alpha N}{2}\log\det\left(\mathbbm{1}-\beta Q\right)\right]\,.
    \end{split}
\end{align}
Finally, after an integration over the remaining Gaussian variables $\mathbf{s}^\mu$, and using \eqref{oracle_properties_general}, we get
\begin{align}
    \mathbb{E}_{(\boldsymbol{\xi}^{\mu})_{\mu>k}}[AB]=\exp\left[-\frac{
    \alpha(1-t)N}{2}\log\det\left(\mathbbm{1}-\beta Q\right)-\frac{1}{2}\sum_{\mu=P(1-t)+1}^P\log\det\left(\mathbbm{1}+\beta Q(v_{\tau^\mu}-1)-(v_{\tau^\mu}-m^2_{\tau^\mu})\beta^2 Q^2\right)\right],
\end{align}
where $\tau^\mu=(1-(\mu-1)/P)$, and $m_{\tau^\mu}=m^\mu$ are the previous retrieval accuracies. It remains to analyze the contribution given by $(\boldsymbol{\xi}^\mu)_{\mu\leq k}$:
\begin{align}
    C:=\exp\left[\frac{\beta N}{2}\sum_{a=1}^n\sum_{\mu=1}^k(m^\mu(\mathbf{x}_a))^2\right]=\int\prod_{a=1}^n\prod_{\mu=1}^k dm^\mu_a\sqrt{\frac{\beta N}{2\pi}}\exp\left[\sum_{a=1}^n\sum_{\mu=1}^k\left(-N\beta\frac{(m^\mu_a)^2}{2}+\beta m^\mu_a\sum_{i=1}^N\xi_i^\mu x_{a,i}\right)\right]\,.
\end{align}

Before plugging the contributions coming from $A$, $B$ and $C$ into $\mathbb{E}\mathcal{Z}_N^n$ we need to introduce a collection of Dirac deltas to fix the desired order parameters, that are organized in the overlap matrix $(Q(\mathbf{x}_a,\mathbf{x}_b))_{a,b=1}^n$:
\begin{align}
    1=\int\prod_{a\leq  b\leq n}dq_{ab}\delta(Q(\mathbf{x}_a,\mathbf{x}_b)-q_{ab})=\int\prod_{a\leq b\leq n}\frac{Ndr_{ab}dq_{ab}}{4\pi i}\exp\left[-\frac{1}{2}\sum_{a,b=1}^nr_{ab}(Nq_{ab}-\sum_ix_{a,i}x_{b,i})\right]\,.
\end{align}
Hence, the averaged replicated partition function, at leading exponential order in $N$, takes the form
\begin{align}
    \begin{split}
        \mathbb{E}\mathcal{Z}_N^n&=\int\prod_{a\leq b\leq n}\frac{Ndr_{ab}dq_{ab}}{4\pi i}\int\prod_{a=1}^n\prod_{\mu=1}^k dm^\mu_a\sqrt{\frac{N \beta}{2\pi}}\exp\left[-\frac{N}{2}\sum_{a,b}r_{ab}q_{ab}-\frac{\beta N}{2}\sum_{a=1}^n\sum_{\mu=1}^k(m^\mu_a)^2 \right]\\
        &\times\exp\left[ -\frac{1}{2}\sum_{\mu=P(1-t)+1}^P\log\det\left(\mathbbm{1}+\beta Q(v_{\tau^\mu}-1)-(v_{\tau^\mu}-m^2_{\tau^\mu})\beta^2 Q^2\right)\right]\\
        &\times\exp\left[-\frac{
        \alpha(1-t)N}{2}\log\det\left(\mathbbm{1}-\beta Q\right)+N\beta^2\Delta\sum_{a\neq b,1}^n\frac{q_{ab}^2}{4}+N\beta\sum_{a=1}^n\Big(-\frac{\lambda}{4}(1-q_{aa})^2+\frac{\beta\Delta-1}{4}q_{aa}^2\Big)\right]\\
        &\times\left(\int \prod_{\mu=1}^kdP_\xi(\xi^\mu)\prod_{a=1}^ndP_\xi(x_{a})\exp\left[\frac{1}{2}\sum_{a,b=1}^nr_{ab}x_{a}x_{b}+\beta\sum_{\mu=1}^k\sum_{a=1}^n{m^\mu_a}\xi^\mu x_{a}\right]\right)^N\, ,
    \end{split}
\end{align}
where we denote $Q=(q_{ab})_{a,b=1}^n$. We can finally express the replicated free entropy with a variational principle coming from a saddle point argument applied to the formula above:
\begin{align}\label{replicated_free_entropy}
    \begin{split}
        &\Phi_n:=\lim_{N\to\infty}\Phi_{N,n}=\frac{1}{n}\text{Extr}\Big\{-\frac{1}{2}\sum_{a,b}r_{ab}q_{ab}
        -\frac{\beta}{2}\sum_{a=1}^n\sum_{\mu=1}^k(m^\mu_a)^2-\frac{
        \alpha(1-t)N}{2}\log\det\left(\mathbbm{1}-\beta Q\right)\\
        &+\beta\sum_{a=1}^n\Big(\frac{\beta\Delta-1}{4}q_{aa}^2-\frac{\lambda}{4}(1-q_{aa})^2\Big)-\frac{\alpha t}{2R}\sum_{\mu=P(1-t)+1}^P\log\det\left[\mathbbm{1}+\beta Q(v_{\tau^\mu}-1)-(v_{\tau^\mu}-m^2_{\tau^\mu})\beta^2 Q^2\right]\\
        &+\beta^2\Delta\sum_{a\neq b,1}^n\frac{q_{ab}^2}{4}+\log\int\prod_{\mu=1}^k\mathbb{E}_{\xi^\mu}\int\prod_{a=1}^n dP_\xi(x_a)\exp\left[\frac{1}{2}\sum_{a,b=1}^nr_{ab}x_{a}x_{b}+\beta\sum_{\mu=1}^k\sum_{a=1}^n m^\mu_a \xi^\mu x_{a}\right]
        \Big\}\,.
    \end{split}
\end{align}
The normalized sum over $\mu=P(1-t)+1,\dots,P$ on the second line can be turned into an integral $\int_0^t\,d\tau\dots$ in the large $N$ limit.
The extremization is taken w.r.t. the collection of parameters $(r_{ab},q_{ab})_{a,b=1}^n$, $(m_a^\mu)_{a=1,\mu=1}^{n,k}$. Within the replica symmetric ansatz
\begin{align}
    \begin{cases}
    r_{ab}=r\,,\quad a\neq b\\
    r_{aa}=-u
    \end{cases}\quad 
    \begin{cases}
    q_{ab}=q\,,\quad a\neq b\\
    q_{aa}=v
    \end{cases}\quad
    m_a^\mu=m^\mu\,,\quad 
    Q=\begin{pmatrix}
    v     &q&q&\dots&q\\
    q     &v&q&\dots&q\\
    q     &q&v&\dots&q\\
    \vdots&\vdots&\vdots&\ddots&\vdots\\
    q&q&q&\dots&v
    \end{pmatrix}\in \mathbb{R}^{n\times n}\,.
\end{align}
The determinants of $\mathbbm{1}-\beta Q$ and $\mathbbm{1}+\beta Q(v_{\tau^\mu}-1)-(v_{\tau^\mu}-m^2_{\tau^\mu})\beta^2 Q^2$ are easily computed:
\begin{align}
    &\det\left(\mathbbm{1}-\beta Q\right)=\left(1-\beta(v-q)\right)^n\left[1-n\frac{\beta q}{1-\beta(v-q)}\right]\\
    \begin{split}
        &\det\left(\mathbbm{1}+\beta Q(v_{\tau^\mu}-1)-(v_{\tau^\mu}-m^2_{\tau^\mu})\beta^2 Q^2\right)=\left[1+\beta(v_{\tau^\mu}-1)(v-q)-(v_{\tau^\mu}-m^2_{\tau^\mu})\beta^2(v-q)^2\right]^{n-1}\\
        &\qquad\times\left[1+\beta(v_{\tau^\mu}-1)(v-q+nq)-(v_{\tau^\mu}-m^2_{\tau^\mu})\beta^2\left(v-q+nq\right)^2\right]\,.
    \end{split}
\end{align}
Further simplifications occur for the other terms in the replicated free entropy. In particular the remaining $\log$ integral is:
\begin{multline}
    \int\prod_{\mu=1}^k\mathbb{E}_{\xi^\mu}\int\prod_{a=1}^n dP_\xi(x_a)\exp\left[\frac{r}{2}\sum_{a\neq b,1}^nx_{a}x_{b}-\frac{u}{2}\sum_{a=1}^nx_a^2+\beta\sum_{\mu=1}^k m^\mu\xi^\mu \sum_{a=1}^n x_{a}\right]=\\
    =\mathbb{E}_Z\int\prod_{\mu=1}^k\mathbb{E}_{\xi^\mu}\prod_{a=1}^n\int dP_\xi(x_a)\exp\left[\sqrt{r}Zx_{a}-\frac{u+r}{2}x_a^2+
    \beta\sum_{\mu=1}^km^\mu\xi^\mu x_{a}\right]=\\
    =\mathbb{E}_Z\mathbb{E}_{\boldsymbol{\xi}}\left[\int dP_\xi(x)\exp\left(\left(Z\sqrt{r}+\beta \mathbf{m}\cdot\boldsymbol{\xi}\right)x-\frac{u+r}{2}x^2\right)\right]^n
\end{multline}where $Z\sim\mathcal{N}(0,1)$, $\boldsymbol{\xi}=(\xi^1,\dots,\xi^k)$, $\mathbf{m}=(m^1,\dots,m^k)$\,. Finally, expanding at first order in $n$ one has:
\begin{align}\label{RS_free_entropy}
    \begin{split}
        &\Phi:=\text{Extr}\Big\{\frac{rq+uv}{2}
        -\beta\sum_{\mu=1}^k\frac{(m^\mu)^2}{2}-\frac{\beta^2\Delta q^2}{4}-\frac{
        \alpha(1-t)}{2}\left[\log\left(1-\beta(v-q)\right)-\frac{\beta q}{1-\beta(v-q)}\right]\\
        &-\frac{\alpha t}{2}\int_0^td\tau\left[\log\left(1+\beta(v_{\tau}-1)(v-q)-(v_{\tau}-m^2_\tau)\beta^2(v-q)^2\right)+ \frac{\beta q(v_\tau-1)-2\beta^2q(v-q)(v_\tau-m^2_\tau)}{1+\beta(v_{\tau}-1)(v-q)-(v_{\tau}-m^2_\tau)\beta^2(v-q)^2}\right]\\
        &\quad+\beta\Big(\frac{\beta\Delta-1}{4}v^2-\frac{\lambda}{4}(1-v)^2\Big)+\mathbb{E}_{Z,\boldsymbol{\xi}}\log\int dP_\xi(x)\exp\left(\left(Z\sqrt{r}+\beta\mathbf{m}\cdot\boldsymbol{\xi}\right)x-\frac{u+r}{2}x^2\right)
        \Big\}\,.
    \end{split}
\end{align}
The correct stationary parameters $v,m,q,u,r$ will be those that maximize the free entropy. Hence it is clear that if $\lambda\to\infty$ we recover the constraint $v=1$.

\subsection{Fixed point equations}
Let us introduce the following notation:
\begin{align}
    \langle\cdot\rangle_{t,\boldsymbol{\xi}}\equiv\langle\cdot\rangle_{t}:=
    \frac{\int dP_\xi(x)\exp\big((Z\sqrt{r}+\beta\mathbf{m}\cdot\boldsymbol{\xi})x-\frac{r+u}{2}x^2\big)(\cdot)}{\int dP_\xi(y)\exp\big((Z\sqrt{r}+\beta\mathbf{m}\cdot\boldsymbol{\xi})y-\frac{r+u}{2}y^2\big)}\,,
\end{align}
where the subscript $t$ emphasizes that we have already reconstructed $R=tP$ patterns.
The stationarity conditions coming from \eqref{RS_free_entropy} are
\begin{align}
    &v=\mathbb{E}_{\boldsymbol{\xi}}\langle X^2\rangle_{t}\\
    \label{m_equation}
    &m^\mu=\mathbb{E}_{\boldsymbol{\xi}}\xi^\mu \langle X\rangle_{t}\,,\quad \mu=1,\dots, k\\
    \label{q_equation}
    &q=\mathbb{E}_{\boldsymbol{\xi}}\langle X\rangle_{t}^2\\
    \begin{split}\label{r_equation}
    &r=\frac{\alpha(1-t)\beta^2 q}{(1-\beta(v-q))^2}+\beta^2\Delta q+
    \alpha t\int_0^t\,d\tau\Big[\frac{2q\beta^2(v_\tau-m^2_\tau)}{1+\beta(v_{\tau}-1)(v-q)-(v_{\tau}-m^2_\tau)\beta^2(v-q)^2}\\
    &\qquad\qquad+q\frac{\beta^2[v_\tau-1-2\beta(v-q)(v_\tau-m_\tau^2)]^2}{[1+\beta(v_{\tau}-1)(v-q)-(v_{\tau}-m^2_\tau)\beta^2(v-q)^2]^2}\Big]
    \end{split}\\
    \begin{split}\label{u_equation}
        &u=\beta\lambda(v-1)+\beta(1-\beta\Delta)v-\alpha(1-t)\beta\frac{1-\beta(v-2q)}{(1-\beta(v-q))^2} -\alpha t\int_0^t\,d\tau\Big[\frac{2v\beta^2(v_\tau-m^2_\tau)-\beta(v_\tau-1)}{1+\beta(v_{\tau}-1)(v-q)-(v_{\tau}-m^2_\tau)\beta^2(v-q)^2}\\
        &\qquad\qquad +q\frac{\beta^2[v_\tau-1-2\beta(v-q)(v_\tau-m^2_\tau)]^2}{[1+\beta(v_{\tau}-1)(v-q)-(v_{\tau}-m^2_\tau)\beta^2(v-q)^2]^2}\Big]\,.
    \end{split}
\end{align}
Notice that the effect of decimation is visible only in the variables $u$ and $r$ that affect the local measure \eqref{local_measure}. 
With a close look to the expression of $r$ we can recognize the three predicted independent noise contribution. The first term is due to pattern interference (noise (b)), and we see that it decreases as $t$ approaches $1$. The second term can be identified with the noise contribution (a), which is due to the original Gaussian noise $\bZ$. The decimation noise contribution (noise (c)) is instead given by the third term, that is expressed in integral form, which correctly takes into account all the history of the process. As anticipated above, the success of decimation is determined by the interplay between noises (b) and (c). Since, as we shall see in Section \ref{sec:numerical_tests}, the retrieval accuracies remain close to one in the range of parameters $\alpha,\Delta$ were the first step of decimation is feasible, the noise contribution (c) will be small. In addition, solving the previous equations for each decimation step shows that the benefit we gain due to the reduction of pattern interference is higher than the penalty we pay for introducing noise with decimation. As a consequence, decimation proves to be a viable strategy for matrix factorization.

For all practical purposes, we will make finite size simulations and use the discretized form present in \eqref{replicated_free_entropy} of the integral accounting for decimation contributions, starting from step $0$, when no pattern has been retrieved yet. Finally, notice that mixed states solutions are possible, with the estimates aligning to more than $1$ pattern, \emph{i.e.} several $m^\mu$'s in \eqref{m_equation} are non-vanishing. This is not desirable in inference, since one wants to estimate one pattern at a time with the best possible performance.

\subsection{Remarks}\label{sanity_checks_section}
First of all, we clarify the relation between our formula and the low-rank formula for the spiked Wigner model. Therefore, let us set $\beta=1/\Delta$, $P=1$, which means $\alpha=0$, and $\lambda=0$. In this case the free entropy reads
\begin{align}
    &\Phi:=\text{Extr}\Big\{\frac{rq+uv}{2}
    -\frac{m^2}{2\Delta}-\frac{ q^2}{4\Delta}+\mathbb{E}_{Z,\boldsymbol{\xi}}\log\int dP_\xi(x)\exp\left(\left(Z\sqrt{r}+\frac{m}{\Delta}\xi\right)x-\frac{u+r}{2}x^2\right)\Big\}
\end{align}
Extremizing w.r.t. $q$ and $v$ we readily find:
\begin{align}
    r=\frac{q}{\Delta}\,,\quad u=0\,.
\end{align}
Plugging this result inside the free entropy yields
\begin{align}
    \Phi:=\text{Extr}\Big\{\frac{q^2}{4\Delta}
    -\frac{m^2}{2\Delta}+\mathbb{E}_{Z,\boldsymbol{\xi}}\log\int dP_\xi(x)\exp\left(\left(Z\sqrt{\frac{q}{\Delta}}+\frac{m\xi}{\Delta}\right)x-\frac{q}{2\Delta}x^2\right)\Big\}\,.
\end{align}
Finally, extremization w.r.t. $q$ and $m$ yields two coupled equations
\begin{align}
    m=\mathbb{E}_\xi\xi\left.\langle X\rangle_t\right|_{r=\frac{q}{\Delta},u=0}\,,\quad q=\mathbb{E}_\xi\left.\langle X\rangle_t^2\right|_{r=\frac{q}{\Delta},u=0}
\end{align}that admit a self consistent solution satisfying a single equation
\begin{align}
    m=q=\mathbb{E}_\xi\xi\left.\langle X\rangle_t\right|_{r=\frac{m}{\Delta},u=0}
\end{align}which is exactly the known fixed point equation for the overlap in the spiked Wigner model.

Secondly, we need to ensure a proper scaling w.r.t. $\beta$. In particular the limit $\lim_{\beta\to\infty}\frac{\Phi}{\beta}$ must be well defined at any decimation step. The only terms in the free entropy that could give rise to overscalings in $\beta$ are
\begin{align}
    \frac{rq+uv}{2}-\frac{\beta^2\Delta q}{4}+\frac{\beta^2\Delta v}{4}\,,\quad\frac{r+u}{2}\,.
\end{align}
The latter in particular appears at the exponent in the gas free entropy in the last line of \eqref{RS_free_entropy}.
Both the fixed point equations for $u$ and $r$ contain terms proportional to $\beta^2$. This issue though is only apparent, and the fixed point remains well defined. To show this let us rewrite the first problematic term as follows:
\begin{align}
    \frac{rq+uv}{2}-\frac{\beta^2\Delta q}{4}+\frac{\beta^2\Delta v}{4}=\frac{-r(v-q)+(u+r)v}{2}+\frac{\beta^2\Delta (v-q)}{4}.
\end{align}
In the limit $\beta\to\infty$ the term
\begin{align}
    -\frac{\beta q}{1-\beta(v-q)}
\end{align}arising from the square bracket in the first line of \eqref{RS_free_entropy} forces $q\to v$ in such a way that $\beta(v-q)<1$ remains of order $O(1)$. Hence $\frac{\beta^2\Delta (v-q)}{4}$ and $r(v-q)=(r/\beta)\beta(v-q)$ are at most of order $O(\beta)$ as they should. It remains to verify that $u+r=O(\beta)$:
\begin{align}
    \begin{split}
        u+r=\beta\lambda(v-1+\beta v)-\beta^2\Delta(v-q)-\frac{\alpha\beta}{1-\beta(v-q)}-\alpha t\int_0^t d\tau \Big[\frac{2\beta^2(v-q)(v_\tau-m^2_\tau)-\beta(v_\tau-1)}{1+\beta(v_{\tau}-1)(v-q)-(v_{\tau}-m^2_\tau)\beta^2(v-q)^2}\Big]\,.
    \end{split}
\end{align}
Again, thanks to the fact that $\beta(v-q)<1$, the correct scaling occurs.

Thirdly, we notice that for Gaussian prior, when patterns are generated from $P_\xi=\mathcal{N}(0,1)$, retrieval is impossible if $\alpha>0$. In fact, from the fixed point equation for $m^\mu$, one can perform a Gaussian integration by parts on the $\xi^\mu$ obtaining:
\begin{align}
    m^\mu=m^\mu\beta\big(   \mathbb{E}\langle X^2\rangle_t-\mathbb{E}\langle X\rangle^2_R   \big)=m^\mu\beta(v-q)
\end{align}which entails $m^\mu=0$ or $\beta(v-q)=1$. The latter though is not possible because it would cause the free entropy to diverge to minus infinity. Hence, the only possibility is to have negligible alignment with all the patterns, $m^\mu=0$. On the contrary if $\alpha=0$, the diverging contribution disappears, and setting $\beta=1/\Delta$ yields the usual PCA estimator overlap $m=q=1-\Delta$.

\section{Low temperature limits}\label{0T_section}

\subsection{Sparse prior}\label{sec:0Tsparse}
Let us express the $\beta\to\infty$ limit of the free entropy with a prior of the form
\begin{align}
    P_\xi=(1-\rho)\delta_0+\frac{\rho}{2}\left[\delta_{-1/\sqrt{\rho}}+\delta_{1/\sqrt{\rho}}\right]\,,\quad \rho\in(0,1)\,.
\end{align}
The case $\rho=1$ shall be discussed separately in the end. For future convenience we introduce the notations
\begin{align}\label{CUrbarr_definition}
    &C:=\beta(v-q)\,\in[0,1)\,,\quad \bar{r}:=r/\beta^2\,,\quad U:=\frac{u+r}{\beta}
\end{align}where $q$ is intended as the stationary value of the overlap solving the fixed point equations. Denote $\mathbf{m}=(m^\mu)_{\mu=1}^k$, where $k$ is the maximum number of condensed patterns. In the low temperature limit the free entropy, re-scaled by $\beta$, and evaluated at the stationary values of the parameters involved has the form
\begin{align}
    \begin{split}
        \frac{1}{\beta}\Phi&=-\frac{\lambda(v-1)^2}{4}-\frac{\bar{r}C}{2}+\frac{Uv}{2}+\frac{\alpha(1-t)v}{2(1-C)}-\frac{v^2}{4}-\frac{\mathbf{m}^2}{2}+\frac{\Delta Cv}{2}+\psi+\frac{\alpha tv}{2}\int_0^t d\tau \frac{2C(v_\tau-m^2_\tau)-(v_\tau-1)}{1
        +(v_\tau-1)C-(v_\tau-m^2_\tau)C^2}
    \end{split}
\end{align}
where
\begin{align}\label{psi_definition}
    \psi=\frac{1}{\beta} \mathbb{E}_{\boldsymbol{\xi},Z}\log\left[1-\rho+\rho\cosh\frac{\beta}{\sqrt{\rho}}\left(Z\sqrt{\bar{r}}+ \mathbf{m}\cdot\boldsymbol{\xi}\right)\exp\left(
   -\frac{\beta U}{2\rho}\right)\right]\,.
\end{align}
When $\beta\to\infty$ we have to distinguish two cases in the $Z$ average:
\begin{align}\label{psi_simplified1}
    \begin{split}
        \psi=O\Big(\frac{1}{\beta}\Big)+&\frac{1}{\beta}\mathbb{E}_{\boldsymbol{\xi}}\left(\int_{-\mathbf{m}\cdot\boldsymbol{\xi}/\sqrt{\bar{r}}+U/2\sqrt{\bar{r}\rho}}^\infty+\int^{-\mathbf{m}\cdot\boldsymbol{\xi}/\sqrt{\bar{r}}-U/2\sqrt{\bar{r}\rho}}_{-\infty}\right)\frac{dz\,e^{-\frac{z^2}{2}}}{\sqrt{2\pi}}
        \log\left[1-\rho+\rho\cosh\frac{\beta}{\sqrt{\rho}}\left(z\sqrt{\bar{r}}+ \mathbf{m}\cdot\boldsymbol{\xi}\right)e^{-\frac{\beta U}{2\rho}}\right].
    \end{split}
\end{align}
The $O(\beta^{-1})$ instead comes from integration on the interval $[-\mathbf{m}\cdot\boldsymbol{\xi}/\sqrt{\bar{r}}-U/2\sqrt{\bar{r}\rho},-\mathbf{m}\cdot\boldsymbol{\xi}/\sqrt{\bar{r}}+U/2\sqrt{\bar{r}\rho}]$ of the same integrand, that can be easily bounded.

Let us now focus on the first integral in \eqref{psi_simplified1}. The hyperbolic cosine and the exponential in $U$ dominate on the other terms in the $\log$. Taking into account the exponential growth in the selected range of $z$-values the first integral can be approximated with:
\begin{multline}
        \mathbb{E}_{\boldsymbol{\xi}}\int_{-\mathbf{m}\cdot\boldsymbol{\xi}/\sqrt{\bar{r}}+U/2\sqrt{\bar{r}\rho}}^\infty\frac{dz}{\sqrt{2\pi}}e^{-\frac{z^2}{2}}\left(\frac{Z\sqrt{\bar{r}}+ \mathbf{m}\cdot\boldsymbol{\xi}}{\sqrt{\rho}}-\frac{U}{2\rho}\right)=
        \sqrt{\frac{\bar{r}}{2\pi\rho}}\mathbb{E}_{\boldsymbol{\xi}}e^{-\frac{1}{2\bar{r}}\left(\frac{U}{2\sqrt{\rho}}-\mathbf{m}\cdot\boldsymbol{\xi}\right)^2}+\\
        +\mathbb{E}_{\boldsymbol{\xi}}\left(\frac{ \mathbf{m}\cdot\boldsymbol{\xi}}{\sqrt{\rho}}-\frac{U}{2\rho}\right)\int_{-\mathbf{m}\cdot\boldsymbol{\xi}/\sqrt{\bar{r}}+U/2\sqrt{\bar{r}\rho}}^\infty\frac{dz}{\sqrt{2\pi}}e^{-\frac{z^2}{2}}\,.
\end{multline}
The second integral in \eqref{psi_simplified1} can be treated similarly. Putting all the terms together one gets
\begin{align}\label{re-scaled_free_entropy1}
    \begin{split}
        \frac{1}{\beta}\Phi&=-\frac{\bar{r}C}{2}+\frac{\Delta Cv}{2}+\frac{Uv}{2}+\frac{\alpha(1-t)v}{2(1-C)}-\frac{v^2+\lambda(v-1)^2}{4}-\frac{\mathbf{m}^2}{2}+\sqrt{\frac{2\bar{r}}{\pi\rho}}\mathbb{E}_{\boldsymbol{\xi}}e^{-\frac{1}{2\bar{r}}\left(\frac{U}{2\sqrt{\rho}}-\mathbf{m}\cdot\boldsymbol{\xi}\right)^2}\\
        &+\mathbb{E}_{\boldsymbol{\xi}}\frac{\mathbf{m}\cdot\boldsymbol{\xi}}{\sqrt{\rho}} \text{erf}\left(\frac{\mathbf{m}\cdot\boldsymbol{\xi}+\frac{U}{2\sqrt{\rho}}}{\sqrt{2\bar{r}}}\right)-\frac{U}{2\rho}\mathbb{E}_{\boldsymbol{\xi}} \left[1
        -\text{erf}\left(\frac{\mathbf{m}\cdot\boldsymbol{\xi}+\frac{U}{2\sqrt{\rho}}}{\sqrt{2\bar{r}}}\right)\right]+\frac{\alpha tv}{2}\int_0^t d\tau \frac{2C(v_\tau-m^2_\tau)-(v_\tau-1)}{1
        +(v_\tau-1)C-(v_\tau-m^2_\tau)C^2}\,.
    \end{split}
\end{align}
Using the fact that all the parameters are evaluated at their stationary values, the previous formula can be further simplified by looking at the limiting version of the fixed point equations. In particular we have that
\begin{align}\label{C_equation}
    C=\sqrt{\frac{2}{\pi\rho\bar{r}}}\mathbb{E}_{\boldsymbol{\xi}}\exp\left(-\left(\frac{U/2\sqrt{\rho}-\mathbf{m}\cdot\boldsymbol{\xi}}{\sqrt{2\bar{r}}}\right)^2\right)\,.
\end{align}
The value of $\bar{r}$ can be found directly from \eqref{r_equation} by multiplying it by $\beta^{-2}$:
\begin{align}
    \label{barr_equation}
    \bar{r}=\frac{\alpha(1-t)v}{(1-C)^2}+\Delta v+\alpha tv\int_0^t\,d\tau\left[\frac{2(v_\tau-m^2_\tau)}{1+(v_\tau-1)C-(v_\tau-m^2_\tau)C^2}+\frac{[v_\tau-1-2C(v_\tau-m^2_\tau)]^2}{[1+(v_\tau-1)C-(v_\tau-m^2_\tau)C^2]^2}\right]
    \,.
\end{align}
Deriving w.r.t. $v$ we get the equation for $U=\frac{u+r}{\beta}$:
\begin{align}
    \label{U_equation_noiseless}
    U=-\Delta C+v+\lambda(v-1)-\frac{\alpha(1-t)}{(1-C)}-\alpha t\int_0^t d\tau \frac{2C(v_\tau-m^2_\tau)-(v_\tau-1)}{1
        +(v_\tau-1)C-(v_\tau-m^2_\tau)C^2}
    \,.
\end{align}
From a derivative w.r.t. $U$ we get an equations for $v$:
\begin{align}
    \label{eq_v_noiseless}
    v=\frac{1}{\rho}\mathbb{E}_{\boldsymbol{\xi}} \left[1
        -\text{erf}\left(\frac{\mathbf{m}\cdot\boldsymbol{\xi}+\frac{U}{2\sqrt{\rho}}}{\sqrt{2\bar{r}}}\right)\right]\,.
\end{align}
We can solve this equation in order to get $U$ as a function of $v$, for instance by dichotomy.
Finally, from \eqref{m_equation} and \eqref{psi_definition}
\begin{align}\label{m_equation_noiseless}
    \mathbf{m}=\mathbb{E}\boldsymbol{\xi}\langle X\rangle_{Z,\boldsymbol{\xi}}=\frac{\partial\psi}{\partial\mathbf{m}}=\mathbb{E}_{\boldsymbol{\xi}}\frac{\boldsymbol{\xi}}{\sqrt{\rho}}\text{erf}\left(\frac{\mathbf{m}\cdot\boldsymbol{\xi}-U/2\sqrt{\rho}}{\sqrt{2\bar{r}}}\right)\,.
\end{align}

If we insert these conditions in \eqref{re-scaled_free_entropy1} we get
\begin{align}
    \label{free_entropy_noiseless}
    \frac{\Phi}{\beta}=\frac{\alpha(1-t)v}{2(1-C)^2}+\Delta Cv-\frac{v^2+\lambda(v-1)^2}{4}+\frac{\mathbf{m}^2}{2}+\frac{\alpha tv}{2}\int_0^t d\tau \frac{4C(v_\tau-m^2_\tau)-(v_\tau-1)[1-(v_\tau-m_\tau^2)C^2]}{[1+(v_\tau-1)C-(v_\tau-m^2_\tau)C^2]^2}\,.
\end{align}
A numerical procedure to find a solution to the previous system of equations is to solve simultaneously \eqref{C_equation} and \eqref{eq_v_noiseless} plugging into them the definitions of $\bar{r}$ and $U$ for a fixed $m$. Then one can iterate \eqref{m_equation_noiseless}. 

Notice that, when $\lambda$ is finite, the problem is not continuous in $\rho=1$, namely sending $\beta\to+\infty$ before or after setting $\rho=1$ is different. This can be seen as a consequence of the non commutation of the two limits $\lim_{\beta\to\infty}$ and $\lim_{\rho\to 1}$ for the quantity $(1-\rho)^{1/\beta}$. In fact, for $\rho=1$ the $O(\beta^{-1})$ contribution in $\psi$ that was discarded before, is no longer negligible. Considering that contribution too would yield a free entropy of the form:
\begin{multline}
        \frac{1}{\beta}\Phi=-\frac{\bar{r}C}{2}+\frac{\Delta Cv}{2}+\frac{Uv}{2}+\frac{\alpha(1-t)v}{2(1-C)}-\frac{v^2+\lambda(v-1)^2}{4}-\frac{\mathbf{m}^2}{2}+\sqrt{\frac{2\bar{r}}{\pi\rho}}\mathbb{E}_{\boldsymbol{\xi}}e^{-\frac{1}{2\bar{r}}\left(\theta(1-\rho)\frac{U}{2\sqrt{\rho}}-\mathbf{m}\cdot\boldsymbol{\xi}\right)^2}\\
        +\mathbb{E}_{\boldsymbol{\xi}}\frac{\mathbf{m}\cdot\boldsymbol{\xi}}{\sqrt{\rho}} \text{erf}\left(\frac{\mathbf{m}\cdot\boldsymbol{\xi}+\theta(1-\rho)\frac{U}{2\sqrt{\rho}}}{\sqrt{2\bar{r}}}\right)-\frac{U}{2\rho}\mathbb{E}_{\boldsymbol{\xi}} \left[1
        -\text{erf}\left(\frac{\mathbf{m}\cdot\boldsymbol{\xi}+\theta(1-\rho)\frac{U}{2\sqrt{\rho}}}{\sqrt{2\bar{r}}}\right)\right]\\
        +\frac{\alpha tv}{2}\int_0^t d\tau \frac{2C(v_\tau-m^2_\tau)-(v_\tau-1)}{1
        +(v_\tau-1)C-(v_\tau-m^2_\tau)C^2}\,,
\end{multline}
where we set $\theta(0)=0$. We see quickly that now, if $\rho=1$, $v=1$ is automatically enforced, whereas it was not so before. This discontinuous behaviour disappears if one sends $\lambda\to+\infty$ from the very beginning, as studied in \cite{MFNN}.

\subsection{Continuous priors}
Consider the same definitions of $\bar{r},C,U$ as above. In this section we deal with priors that are symmetric and absolutely continuous over the Lebesgue measure, with density $p(x)$. We require the density to be finite at the boundaries of the support $[-a,a]$, or to go to zero with at most polynomial speed, and to be non-vanishing in the interior of the support. An example is the uniform distribution over $[-\sqrt{3},\sqrt{3}]$. The prior dependent part in the free entropy is still
\begin{align}
    \psi:=\frac{1}{\beta}\mathbb{E}_{Z,\boldsymbol{\xi}}\log\int dP_\xi(x)e^{\beta(Z\sqrt{\bar{r}}+\mathbf{m}\cdot\boldsymbol{\xi})x-\frac{\beta U}{2}x^2}\,.
\end{align}
We separate the quenched Gaussian integral from the expectation w.r.t. $\boldsymbol{\xi}$, and we perform the following changes of variables: $z\mapsto z/\sqrt{\bar{r}}$, $z\mapsto z-\mathbf{m}\cdot\boldsymbol{\xi}$. This yields
\begin{multline}
    \psi=\frac{1}{\beta}\mathbb{E}_{\boldsymbol{\xi}}
    \int\frac{dz}{\sqrt{2\pi\bar{r}}}e^{-\frac{(z-\mathbf{m}\cdot\boldsymbol{\xi})^2}{2\bar{r}}}\log\int_{-a}^a dx 
    p(x)e^{-\frac{\beta 
    U}{2}\left(x-\frac{z}{U}\right)^2+\frac{\beta z^2}{2U}}=\\
    =\frac{\bar{r}+\mathbf{m}^2}{2U}+\frac{1}{\beta}\mathbb{E}_{\boldsymbol{\xi}}
    \int\frac{dz}{\sqrt{2\pi\bar{r}}}e^{-\frac{(z-\mathbf{m}
    \cdot\boldsymbol{\xi})^2}{2\bar{r}}}\log\int_{-a}^a dx 
    p(x)e^{-\frac{\beta 
    U}{2}\left(x-\frac{z}{U}\right)^2}=:\frac{\bar{r}+\mathbf{m}^2}{2U}+\bar{\psi}\,.
\end{multline}
The integral inside the logarithm in $\bar{\psi}$ can be computed by Laplace's approximation when $\beta$ is large. However, the location of the maximum of the exponent depends on the value of $z$. In particular if $z\in[-Ua,Ua]$ then the maximum point falls inside the support of $p(x)$. Otherwise, given the quadratic nature of the exponent, the maximum in $x$ will be attained at the boundaries of the support $-a$ and $a$.
Hence the $z$-integral must be divided into three segments. Let us first consider:
\begin{align}
    \text{I}=\frac{1}{\beta}\mathbb{E}_{\boldsymbol{\xi}}
    \int_{-Ua}^{Ua}\frac{dz}{\sqrt{2\pi\bar{r}}}e^{-\frac{(z-\mathbf{m}\cdot\boldsymbol{\xi})^2}{2\bar{r}}}\log\int_{-a}^a dx 
    p(x)e^{-\frac{\beta U}{2}\left(x-\frac{z}{U}\right)^2}\xrightarrow[]{\beta\to\infty}0
\end{align}
because the exponent equals $0$ at the maximum. Hence no exponential contribution in $\beta$ is given, that is able to constrast the $1/\beta$ in front. 

Let us turn to a second contribution:
\begin{align}
    \begin{split}
    \text{II}&=\frac{1}{\beta}\mathbb{E}_{\boldsymbol{\xi}}
    \int_{Ua}^{+\infty}\frac{dz}{\sqrt{2\pi\bar{r}}}e^{-\frac{(z-\mathbf{m}\cdot\boldsymbol{\xi})^2}{2\bar{r}}}
    \log\int_{-a}^a dx  p(x)e^{-\frac{\beta U}{2}\left(x-\frac{z}{U}\right)^2}\xrightarrow[]{\beta\to\infty}
    -\frac{U}{2}\mathbb{E}_{\boldsymbol{\xi}}
    \int_{Ua}^{+\infty}\frac{dz}{\sqrt{2\pi\bar{r}}}e^{-\frac{(z-\mathbf{m}\cdot\boldsymbol{\xi})^2}{2\bar{r}}}\left(a-\frac{z}{U}\right)^2
    \end{split}
\end{align}
From the square in the integrand we get three sub-contributions.
\begin{align}
    \text{IIA}=-\frac{Ua^2}{2}\mathbb{E}_{\boldsymbol{\xi}}
    \int_{Ua}^{+\infty}\frac{dz}{\sqrt{2\pi\bar{r}}}e^{-\frac{(z-\mathbf{m}\cdot\boldsymbol{\xi})^2}{2\bar{r}}}=-\frac{Ua^2}{4}\text{erfc}\Big(\frac{Ua-\mathbf{m}\cdot\boldsymbol{\xi}}{\sqrt{2\bar{r}}}\Big)
\end{align}
where the last step follows from a simple change of variables.
The second one, with a shift in the integration variable, is
\begin{align}
    \text{IIB}=a\mathbb{E}_{\boldsymbol{\xi}}
    \int_{Ua-\mathbf{m}\cdot\boldsymbol{\xi}}^{+\infty}\frac{dz}{\sqrt{2\pi\bar{r}}}
    e^{-\frac{z^2}{2\bar{r}}}(z+\mathbf{m}\cdot\boldsymbol{\xi})=a\sqrt{\frac{\bar{r}}{2\pi}}\mathbb{E}_{\boldsymbol{\xi}}e^{-\frac{(Ua-\mathbf{m}\cdot\boldsymbol{\xi})^2}{2\bar{r}}}+a\mathbb{E}_{\boldsymbol{\xi}}
    \mathbf{m}\cdot\boldsymbol{\xi}\,\text{erfc}\Big(\frac{Ua-\mathbf{m}\cdot\boldsymbol{\xi}}{\sqrt{2\bar{r}}}\Big)\,.
\end{align}
Finally, with the same shift in the integration variable, we get a third contribution:
\begin{multline}
    \text{IIC}=-\frac{1}{2U}\mathbb{E}_{\boldsymbol{\xi}}
    \int_{Ua-\mathbf{m}\cdot\boldsymbol{\xi}}^{+\infty}\frac{dz}{\sqrt{2\pi\bar{r}}}
    e^{-\frac{z^2}{2\bar{r}}}(z^2+2z\mathbf{m}\cdot\boldsymbol{\xi}+(\mathbf{m}\cdot\boldsymbol{\xi})^2)=-\frac{1}{2U}\sqrt{\frac{\bar{r}}{2\pi}}\mathbb{E}_{\boldsymbol{\xi}}(Ua+\mathbf{m}\cdot\boldsymbol{\xi})e^{-\frac{(Ua-\mathbf{m}\cdot\boldsymbol{\xi})^2}{2\bar{r}}}\\
    -\frac{1}{4U}\mathbb{E}_{\boldsymbol{\xi}}
    (\mathbf{m}\cdot\boldsymbol{\xi})^2\,\text{erfc}\Big(\frac{Ua-\mathbf{m}\cdot\boldsymbol{\xi}}{\sqrt{2\bar{r}}}\Big)
    -\frac{\bar{r}}{4U}\mathbb{E}_{\boldsymbol{\xi}}
    \text{erfc}\Big(\frac{Ua-\mathbf{m}\cdot\boldsymbol{\xi}}{\sqrt{2\bar{r}}}\Big)\,.
\end{multline}

Now, it remains to compute the last gaussian integral:
\begin{align}
\begin{split}
    \text{III}&=\frac{1}{\beta}\mathbb{E}_{\boldsymbol{\xi}}
    \int_{-\infty}^{Ua}\frac{dz}{\sqrt{2\pi\bar{r}}}e^{-\frac{(z-\mathbf{m}\cdot\boldsymbol{\xi})^2}{2\bar{r}}}
    \log\int_{-a}^a dx  p(x)e^{-\frac{\beta U}{2}\left(x-\frac{z}{U}\right)^2}\,.
    \end{split}    
\end{align}
Thanks to the parity of $p(x)$, if we perform the changes of variables $z\mapsto-z$, $\boldsymbol{\xi}\mapsto-\boldsymbol{\xi}$, $x\mapsto-x$ we find that II$=$III. Hence we can finally recompose $\psi$:
\begin{align}
\begin{split}
    \psi=\frac{\bar{r}+\mathbf{m}^2}{2U}+2\text{II}=-\frac{Ua^2}{2}+\frac{1}{U}\sqrt{\frac{\bar{r}}{2\pi}}\mathbb{E}_{\boldsymbol{\xi}}(Ua-\mathbf{m}\cdot\boldsymbol{\xi})e^{-\frac{(Ua-\mathbf{m}\cdot\boldsymbol{\xi})^2}{2\bar{r}}}+\mathbb{E}_{\boldsymbol{\xi}}\frac{\bar{r}+(Ua-\mathbf{m}\cdot\boldsymbol{\xi})^2}{2U}\text{erf}\Big(\frac{Ua-\mathbf{m}\cdot\boldsymbol{\xi}}{\sqrt{2\bar{r}}}\Big)\,.
\end{split}
\end{align}
and the final form of the asymptotic free entropy is
\begin{multline}
    \frac{\Phi}{\beta}\xrightarrow[]{\beta\to\infty}-\frac{\bar{r}C}{2}+\frac{U(v-a^2)}{2}-\frac{\mathbf{m}^2}{2}+\frac{\alpha(1-t)v}{2(1-C)}+\frac{\Delta Cv}{2}-\frac{v^2+\lambda(v-1)^2}{4}
    +\frac{1}{U}\sqrt{\frac{\bar{r}}{2\pi}}\mathbb{E}_{\boldsymbol{\xi}}(Ua-\mathbf{m}\cdot\boldsymbol{\xi})e^{-\frac{(Ua-\mathbf{m}\cdot\boldsymbol{\xi})^2}{2\bar{r}}}\\+
    \mathbb{E}_{\boldsymbol{\xi}}\frac{\bar{r}+(Ua-\mathbf{m}\cdot\boldsymbol{\xi})^2}{2U}\text{erf}\Big(\frac{Ua-\mathbf{m}\cdot\boldsymbol{\xi}}{\sqrt{2\bar{r}}}\Big)+\frac{\alpha tv}{2}\int_0^t d\tau \frac{2C(v_\tau-m^2_\tau)-(v_\tau-1)}{1+(v_\tau-1)C-(v_\tau-m^2_\tau)C^2}\,.
\end{multline}

The saddle point equations can be obtained by deriving the previous formula. The gradient w.r.t. $\mathbf{m}$ yields:
\begin{align}\label{eq:m_uniform_prior}
    \mathbf{m}=\mathbb{E}_{\boldsymbol{\xi}}\frac{\boldsymbol{\xi}}{U}\Big[-\sqrt{\frac{2\bar{r}}{\pi}}e^{-\frac{(Ua-\mathbf{m}\cdot\boldsymbol{\xi})^2}{2\bar{r}}}+
    (Ua-\mathbf{m}\cdot\boldsymbol{\xi})\text{erf}\Big(\frac{\mathbf{m}\cdot\boldsymbol{\xi}-Ua}{\sqrt{2\bar{r}}}\Big)
    \Big]\,.
\end{align}
The derivative w.r.t. $\bar{r}$ gives the equation for $C$:
\begin{align}\label{eq:C_uniform_prior}
    C=\frac{1}{U}\mathbb{E}_{\boldsymbol{\xi}}\text{erf}\Big(\frac{Ua-\mathbf{m}\cdot\boldsymbol{\xi}}{\sqrt{2\bar{r}}}\Big)\,.
\end{align}
Deriving w.r.t. $U$ yields an equation for $v$:
\begin{align}\label{eq:v_uniform_prior}
    \frac{a^2-v}{2}=\frac{1}{U^2}\sqrt{\frac{\bar{r}}{2\pi}}\mathbb{E}_{\boldsymbol{\xi}}(Ua+\mathbf{m}\cdot\boldsymbol{\xi})e^{-\frac{(Ua-\mathbf{m}\cdot\boldsymbol{\xi})^2}{2\bar{r}}}-\mathbb{E}_{\boldsymbol{\xi}}\Big[
    \frac{\bar{r}+(Ua-\mathbf{m}\cdot\boldsymbol{\xi})^2}{2U^2}-\frac{a}{U}(Ua-\mathbf{m}\cdot\boldsymbol{\xi})
    \Big]\text{erf}\Big(\frac{Ua-\mathbf{m}\cdot\boldsymbol{\xi}  }{\sqrt{2\bar{r}}}\Big)\,.
\end{align}
In all the previous equations $\bar{r}$ and $U$ must be considered as the following functions:
\begin{align}\label{eq:rbar_uniform_prior}
    \bar{r}&=\frac{\alpha(1-t) v}{(1-C)^2}+\Delta v+\alpha tv\int_0^t\,d\tau\left[\frac{2(v_\tau-m^2_\tau)}{1+(v_\tau-1)C-(v_\tau-m^2_\tau)C^2}+\frac{[v_\tau-1-2C(v_\tau-m^2_\tau)]^2}{[1+(v_\tau-1)C-(v_\tau-m^2_\tau)C^2]^2}\right]\\
    U&=-\Delta C+v+\lambda(v-1)-\frac{\alpha(1-t)}{(1-C)}-\alpha t\int_0^t d\tau \frac{2C(v_\tau-m^2_\tau)-(v_\tau-1)}{1
        +(v_\tau-1)C-(v_\tau-m^2_\tau)C^2}\,.
\end{align}
Equations \eqref{eq:C_uniform_prior} and \eqref{eq:v_uniform_prior} shall be solved simultaneously at any iteration step for $\mathbf{m}$. This will yield a convergent algorithm to solve the system of equations.

To evaluate the free entropy at the solution of the previous system of saddle point equations we first enforce equation \eqref{eq:v_uniform_prior}, obtaining:
\begin{multline}
    \frac{\Phi}{\beta}\xrightarrow[]{\beta\to\infty}-\frac{\bar{r}C}{2}+\frac{U(v-a^2)}{2}-\frac{\mathbf{m}^2}{2}+\frac{\alpha(1-t)v}{2(1-C)}+\frac{\Delta Cv}{2}-\frac{v^2+\lambda(v-1)^2}{4}
    +\frac{1}{U}\sqrt{\frac{\bar{r}}{2\pi}}\mathbb{E}_{\boldsymbol{\xi}}(Ua-\mathbf{m}\cdot\boldsymbol{\xi})e^{-\frac{(Ua-\mathbf{m}\cdot\boldsymbol{\xi})^2}{2\bar{r}}}\\+
    \mathbb{E}_{\boldsymbol{\xi}}\frac{\bar{r}+(Ua-\mathbf{m}\cdot\boldsymbol{\xi})^2}{2U}\text{erf}\Big(\frac{Ua-\mathbf{m}\cdot\boldsymbol{\xi}}{\sqrt{2\bar{r}}}\Big)+\frac{\alpha tv}{2}\int_0^t d\tau \frac{2C(v_\tau-m^2_\tau)-(v_\tau-1)}{1+(v_\tau-1)C-(v_\tau-m^2_\tau)C^2}\,.
\end{multline}
Using the equation for $C$ \eqref{eq:C_uniform_prior} we see that the first term in the first line and the first term in the second line can be summed together. After some algebra, imposing also \eqref{eq:m_uniform_prior} we get
\begin{align}
    \frac{\Phi}{\beta}\xrightarrow[]{\beta\to\infty}\frac{\bar{r}C}{2}+\frac{\mathbf{m}^2}{2}+\frac{\alpha(1-t)v}{2(1-C)}+\frac{\Delta Cv}{2}-\frac{v^2+\lambda(v-1)^2}{4}+\frac{\alpha tv}{2}\int_0^t d\tau \frac{2C(v_\tau-m^2_\tau)-(v_\tau-1)}{1+(v_\tau-1)C-(v_\tau-m^2_\tau)C^2}\,.
\end{align}
Finally, inserting also \eqref{eq:rbar_uniform_prior} we get
\begin{align}
    \frac{\Phi}{\beta}=\frac{\alpha(1-t)v}{2(1-C)^2}+\Delta Cv-\frac{v^2+\lambda(v-1)^2}{4}+\frac{\mathbf{m}^2}{2}+\frac{\alpha tv}{2}\int_0^t d\tau \frac{4C(v_\tau-m^2_\tau)-(v_\tau-1)[1-(v_\tau-m_\tau^2)C^2]}{[1+(v_\tau-1)C-(v_\tau-m^2_\tau)C^2]^2}\,.
\end{align}
which surprisingly coincides with \eqref{free_entropy_noiseless}.

\section{Phase diagrams for the first decimation step}\label{sec:diagrams}
The starting point of the decimation process is of crucial importance for its success. In fact, if we were to subtract an estimate $\boldsymbol{\eta}\boldsymbol{\eta}^\intercal/\sqrt{N}$ from the observations $\boldsymbol{Y}$ where $\boldsymbol{\eta}$ had a negligible alignment with all the  patterns, we would actually introducing further noise without decreasing the rank of the hidden matrix: decimation would be bound to fail.

At the $1$-st step ($R=0$ or $t=0$) the replica symmetric decimation free entropy is simply that of a Hopfield model with Gaussian noise:
\begin{align}
    \label{0th-step_RS_free_entropy}
    &\Phi(t=0):=\text{Extr}\Big\{\frac{rq+uv}{2}
    -\beta\sum_{\mu=1}^k\frac{(m^\mu)^2}{2}-\frac{\beta^2\Delta q^2}{4}-\frac{
    \alpha}{2}\left[\log\left(1-\beta(v-q)\right)-\frac{\beta q}{1-\beta(v-q)}\right]\\
    &\quad+\beta\Big(\frac{\beta\Delta-1}{4}v^2-\frac{\lambda}{4}(1-v)^2\Big)+\mathbb{E}_{Z,\boldsymbol{\xi}}\log\int dP_\xi(x)\exp\left(\left(Z\sqrt{r}+\beta\mathbf{m}\cdot\boldsymbol{\xi}\right)x-\frac{u+r}{2}x^2\right)
    \Big\}\,.
\end{align}
The set of fixed point equations then simplifies remarkably to
\begin{align}
    &v=\mathbb{E}_{\boldsymbol{\xi}}\langle X^2\rangle_{t}\,,\quad m^\mu=\mathbb{E}_{\xi}\xi \langle X\rangle_{t}\,,\quad q=\mathbb{E}_{\boldsymbol{\xi}}\langle X\rangle_{t}^2\\
    &r=\frac{\alpha\beta^2 q}{(1-\beta(v-q))^2}+\beta^2\Delta q\,,\quad u=\beta\lambda(v-1)+\beta(1-\beta\Delta)v-\alpha\beta\frac{1-\beta(v-2q)}{(1-\beta(v-q))^2}\,.
\end{align}
where we have assumed condensation onto only one pattern. 

\begin{figure}[t]
    \centering
    \includegraphics[width=0.48\textwidth]{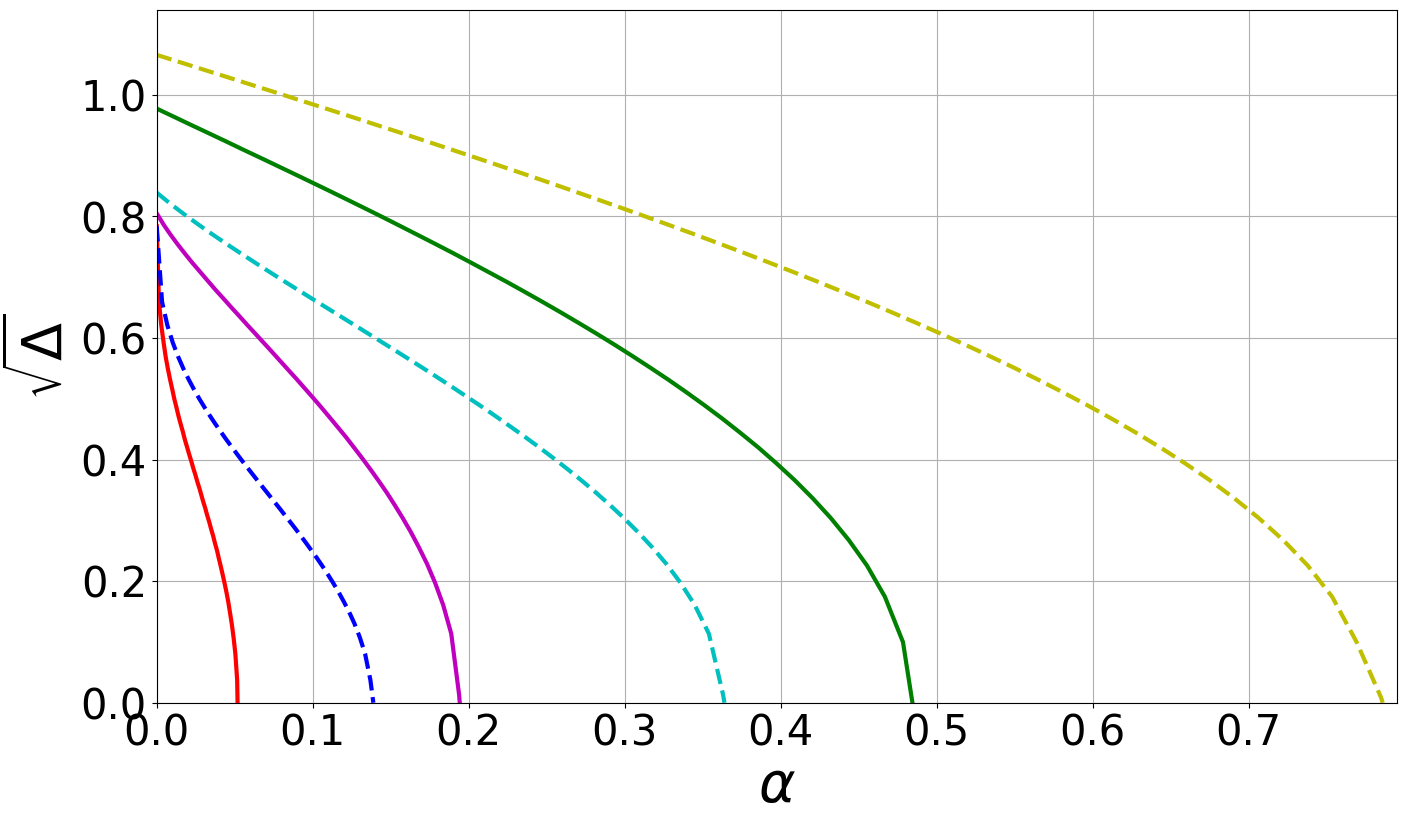}
    \includegraphics[width=0.4875\textwidth]{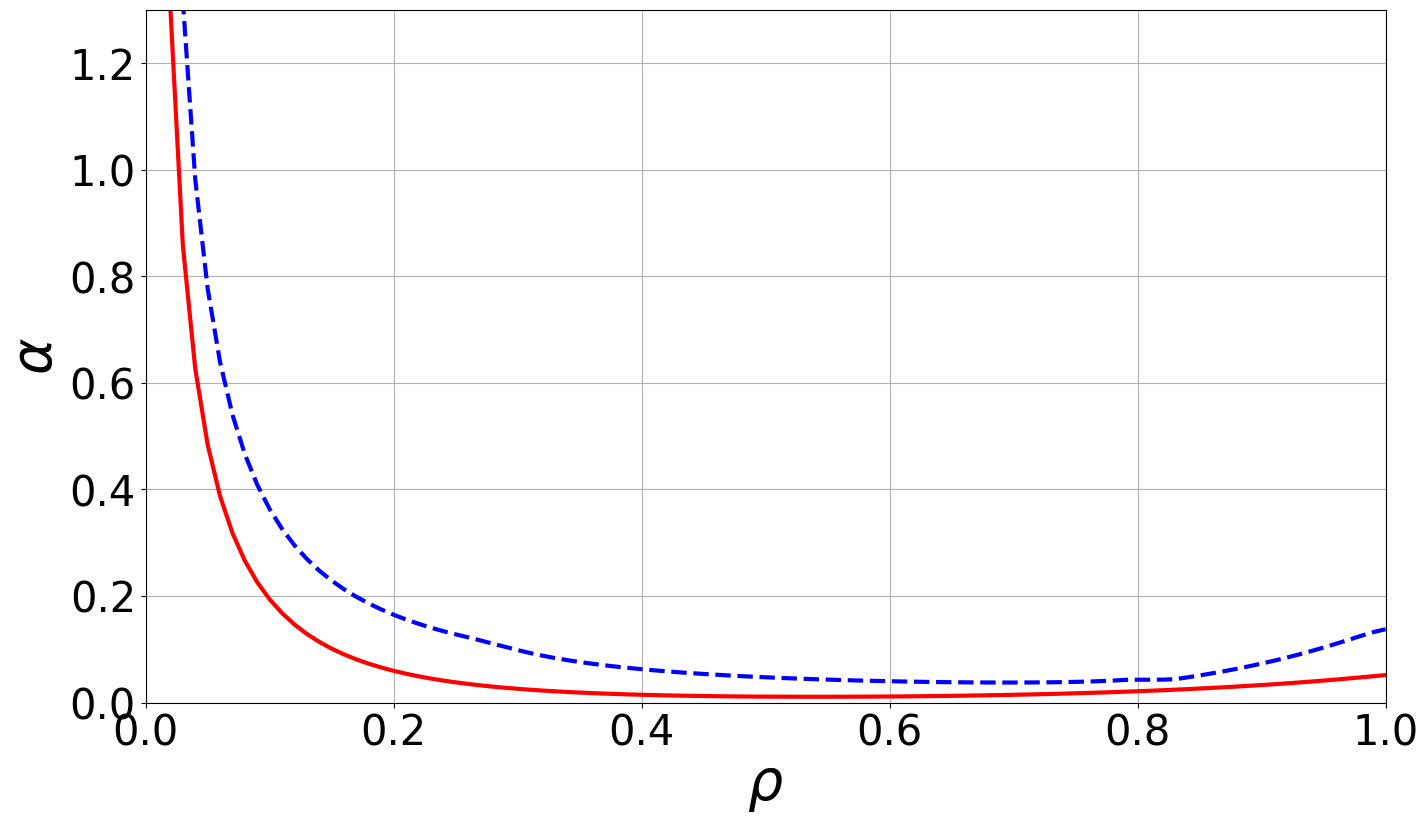}
    \caption{\small \textbf{Left panel}: 
    Phase diagram for the first step of decimation in the case of sparse Ising prior. The lines show the
    zero temperature phase diagram for different values of the sparsity parameter $\rho$ (using $\lambda\to\infty$). Dashed lines plot the storage capacity as a function of $\Delta$. Solid lines signal the thermodynamic transition from the glassy phase to the retrieval phase, when configurations with non vanishing magnetizations with the patterns become thermodynamically stable. The blue and red lines are for $\rho=1$; cyan and magenta for $\rho=0.1$; green and yellow for $\rho=0.05$. \textbf{Right panel}: zero temperature storage capacity $\alpha_c$ and critical thermodynamic storage $\alpha_F$, in dashed blue and solid red lines respectively, versus sparsity $\rho$ in the case $\Delta=0$ (using $\lambda\to\infty$). This plot tracks the behaviour of the intersection of the dashed and solid lines with the $x$-axis in the left panel as $\rho$ varies in $(0,1]$.}
    \label{fig:phasediag_sparse}
\end{figure}

Starting from these equations, one can specialize to the different $0$ temperature limits that exhibit interesting features. For instance in the left panel of \figurename\,\ref{fig:phasediag_sparse}, we see how the phase diagram at $0$ temperature changes as sparsity increases when $\lambda\to\infty$ for the sparse Ising prior. It appears that sparsity increases the retrival region and also the storage capacity. From the right panel we indeed see that the critical storage capacity in the noiseless limit $\Delta=0$ diverges when $\rho\to 0$. This observation can be turned into an analytical statement as follows. To begin with, we notice that
\begin{align}
    C=\frac{2(1-\rho)}{\sqrt{2\pi\bar{r}\rho}} e^{-\frac{U^2}{8\bar{r}\rho}}+\frac{\rho}{\sqrt{2\pi\bar{r}\rho}}\left[e^{-\left(\frac{U/2+m}{\sqrt{2\bar{r}\rho}}\right)^2}+e^{-\left(\frac{U/2-m}{\sqrt{2\bar{r}\rho}}\right)^2}\right]\xrightarrow[]{\rho\to0}0\,,
\end{align}
exponentially fast, and
\begin{align}
    \bar{r}\xrightarrow[]{\rho\to0}v(\alpha+\Delta)\,.
\end{align}As a consequence the equation \eqref{U_equation_noiseless} for $U$ reduces to:
\begin{align}
    U=v+\lambda(v-1)-\alpha\quad\Rightarrow\quad v=\frac{U+\alpha+\lambda}{\lambda+1}\,.
\end{align}
We argue that $U$ is always positive, as it serves as a norm regulator on the estimator, and we verified this statement numerically. This implies that $v$ is always strictly positive. Equation \eqref{eq_v_noiseless} can thus be rewritten as an equation for $U$ that reads as:
\begin{align}
    \frac{U+\alpha+\lambda}{\lambda+1}=\frac{1}{\rho}-\frac{1-\rho}{\rho}\text{erf}\Big(\frac{U}{2\sqrt{2\rho\bar{r}}}\Big)-\frac{1}{2}\Big[\text{erf}\Big(\frac{U/2-m}{\sqrt{2\bar{r}\rho}}\Big)+\text{erf}\Big(\frac{U/2+m}{\sqrt{2\bar{r}\rho}}\Big)\Big]\,.
\end{align}
The error function saturates exponentially fast to $1$ when $\rho\to0$, and this entails
\begin{align}
    \frac{U+\alpha+\lambda}{\lambda+1}=1-\frac{1}{2}\Big[\text{erf}\Big(\frac{U/2-m}{\sqrt{2\bar{r}\rho}}\Big)+\text{erf}\Big(\frac{U/2+m}{\sqrt{2\bar{r}\rho}}\Big)\Big]+O\big(e^{-K/\rho}\big)
\end{align}
for some positive constant $K$, and up to logarithmic corrections at the exponent in the remainder. The argument in the square brackets can go either to $0$ or to $2$ depending on the signs of the arguments in the error functions. However, the second possibility, that would correspond to $U/2>|m|$, is not possible, since the l.h.s. cannot converge to $0$ thanks to the positivity of $U$. Hence, the only alternative we have is that $U/2<|m|$, which is also verified numerically. This implies that the limiting equation for $\rho\to0$ appears as
\begin{align}
    \frac{U+\alpha+\lambda}{\lambda+1}=1\quad\Rightarrow\quad \lim_{\rho\to 0}U=1-\alpha\quad\Rightarrow\quad\lim_{\rho\to 0}v=1\,.
\end{align}
Finally, using the condition $U/2<|m|$, the limit of the magnetization can be easily computed from \eqref{m_equation_noiseless}:
\begin{align}
    m=\frac{1}{2}\Big[\text{erf}\Big(\frac{m-U/2}{\sqrt{2\bar{r}\rho}}\Big)+\text{erf}\Big(\frac{U/2+m}{\sqrt{2\bar{r}\rho}}\Big)\Big]\xrightarrow[]{\rho\to0}1\,.
\end{align} The behaviour depicted so far of the variables $m,C,v,\bar{r}$ and $U$ has been verified numerically for various values of $\lambda$, $\alpha$ and $\Delta$. 
\begin{figure}[ht]
    \centering
    \includegraphics[width=.6\textwidth]{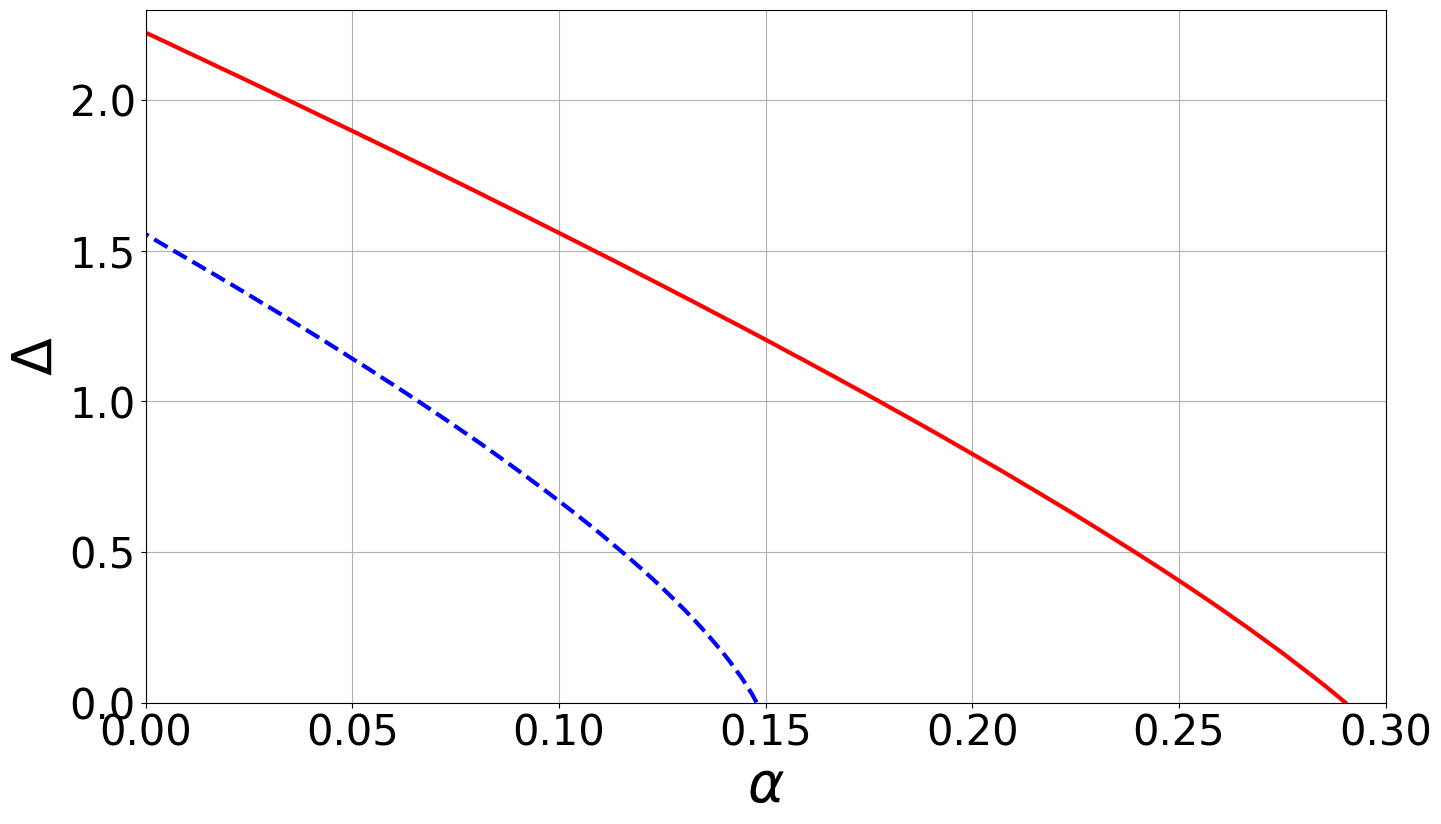}
    \caption{\small Zero temperature phase diagram for uniform prior supported on $[-\sqrt{3},\sqrt{3}]$ and $\lambda=0$. The solid line represents the thermodynamic phase transition. Below it, probability is dominated by those 'retrieval' states that have a non vanishing Mattis magnetization with one pattern. The dashed blue line represents a performance transition: below it the mean configuration of the Boltzmann-Gibbs measure has a better performance in reconstructing the pattern than the null estimator $\boldsymbol{\eta}_{null}=0$.}
    \label{fig:0T_phasediag_uniform}
\end{figure}

In \figurename\,\ref{fig:0T_phasediag_uniform} we plot the phase diagram for a continuous uniform prior supported on $[-\sqrt{3},\sqrt{3}]$ with $\lambda=0$. We verified that once that a magnetization $m\neq 0$ is a solution to the fixed point equations, then it is also thermodynamically stable, namely its free entropy is automatically bigger than that of the $m=0$ solution, contrary to what happens for the discrete priors discussed above. The dashed line here does not signal a proper phase transition, but it is the location of the phase space where the mean square error in the reconstruction of the single pattern outperforms the null estimator $\boldsymbol{\eta}_{null}=0$, namely when:
\begin{align}
    \text{MSE}(\boldsymbol{\eta};\boldsymbol{\xi})=\frac{1}{N}\Vert \boldsymbol{\xi}-\langle\boldsymbol{\eta}\rangle\Vert^2\simeq 1+v-2m<1\,,
\end{align}
where the approximate equality holds true in the $N\to\infty$ and $\beta\to\infty$ limit. Notice that the performance of a Bayes-optimal estimator is always upper bounded by $1$ thanks to the Nishimori identities, hence it is always at least as good as the null estimator.

\section{Numerical tests}\label{sec:numerical_tests}

\subsection{Testing the saddle point equations with AMP}
In order to test our theoretical predictions, we need an algorithm that is able to sample from the Botlzmann-Gibbs measure, or at least that can estimate its marginals, namely the local magnetizations. Approximate message passing is an algorithm that serves the purpose. Furthermore, one needs to integrate the decimation scheme into it. The resulting algorithm was called \emph{decimated AMP} (see Algorithm \ref{alg:decimated-AMP}), which first appeared informally in \cite{camilli2023new}, and then refined in \cite{Jean_sublinear}.

It is possible to derive a suitable AMP from the set of belief propagation equations for the Boltzmann-Gibbs measure:
\begin{align}
    &\hat{m}^t_{(ij)\to i}(x_i)\propto \int dx_j \hat{m}^t_{j\to(ij)}(x_j)\exp\Big[\frac{\beta}{\sqrt{N}}Y_{ij}x_ix_j-\frac{\beta(1+\lambda)}{2N}x_i^2x_j^2\Big]\\
    &{m}^{t+1}_{i\to(ij)}(x_i)\propto dP_\xi(x_i)\exp\Big(\frac{\beta\lambda x_i^2}{2}\Big)\prod_{k\neq i,j}\hat{m}^t_{(ki)\to i}(x_i)\,,
\end{align}
by expanding in $N$ and keeping the leading order. The resulting algorithm, which takes as input an appropriate initialization and the data, reads:
\begin{align}
    &\mathbf{x}^{t+1}=f(\mathbf{A}^t,\mathbf{B}^t)\,, \quad \mathbf{v}^{t+1}=\partial_a f(\mathbf{A}^t,\mathbf{B}^t)\\
    \label{DAMP:cavityfieldA}
    &\mathbf{A}^t=\frac{\beta}{\sqrt{N}}\mathbf{Y}\mathbf{x}^t-\frac{\beta^2}{N}\mathbf{x}^{t-1}\circ (\mathbf{Y}^{\circ2}\mathbf{v}^t)\\
    \label{DAMP:cavityfieldB}
    &\mathbf{B}^t=\frac{\beta}{N}\big((1-\mathbf{Y}^{\circ 2})\mathbf{v}+\Vert\mathbf{x}^t\Vert^2\big)+\frac{\beta\lambda}{N}\sum_{i=1}^N\big(v_i^t+(x_i^t)^2-1\big)
\end{align}
where constants are summed element/component-wise, $\circ$ is the Hadamard entry-wise product (or power), and as denoisers we have chosen the local means
\begin{align}
    f(a,b)=\frac{\int dP_\xi(x)x\exp(ax-\frac{bx^2}{2})}{\int dP_\xi(y)\exp(ay-\frac{by^2}{2})}
\end{align}
that are also applied component-wise to vectors. We denote this algorithm in a compact way by $\text{AMP}(\mathbf{Y},\mathbf{x}^0,\mathbf{v}^0)$, and it is run until the marginals stabilize with a certain tolerance. The above AMP is used to estimate the first and second moment marginals of the Boltzmann-Gibbs measure: $x_i^\infty\simeq \langle x_i\rangle$, $v_i^\infty\simeq \langle x_i^2\rangle-\langle x_i\rangle^2$. Of course the very same algorithm can be run on the set of modified observations $\mathbf{Y}_R$ in \eqref{eq:modified_obs}, which is accessible to the statistician at every decimation step. 

\begin{figure}[t]
    \centering
    \includegraphics[width=0.325\textwidth]{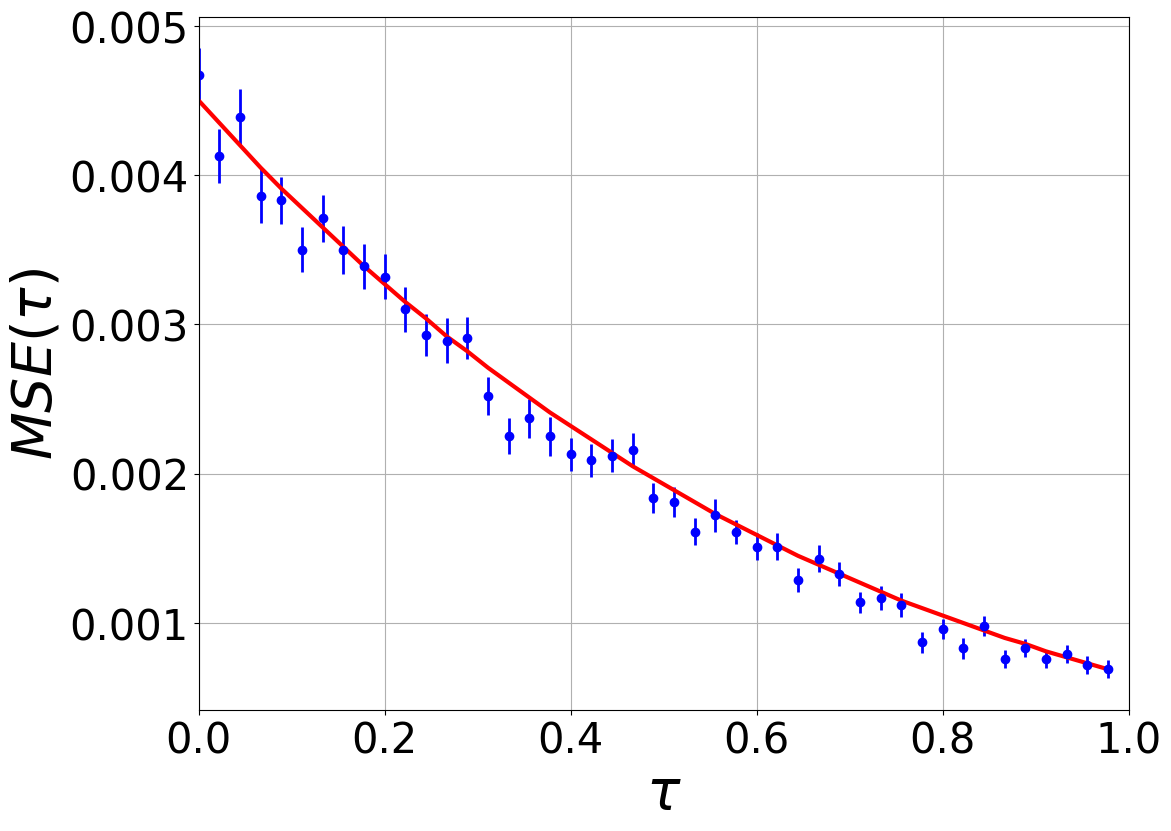}
    \includegraphics[width=0.325\textwidth]{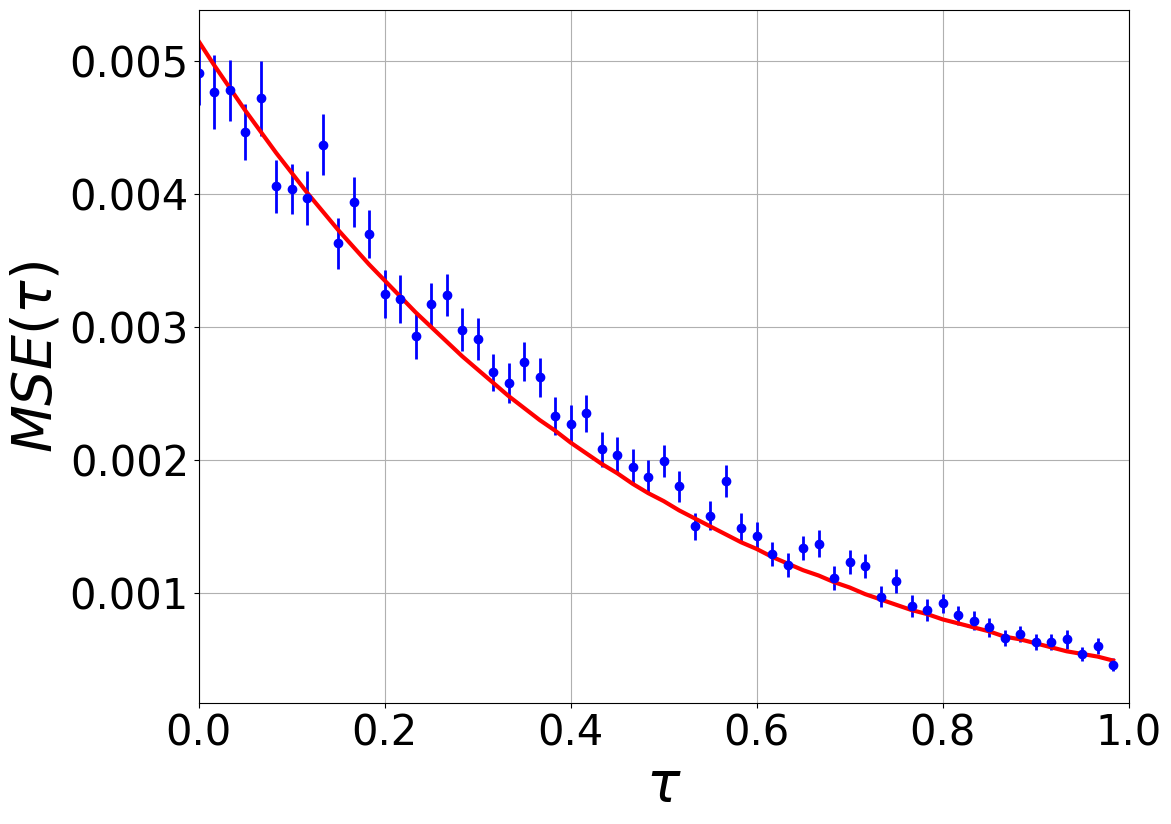}
    \includegraphics[width=0.325\textwidth]{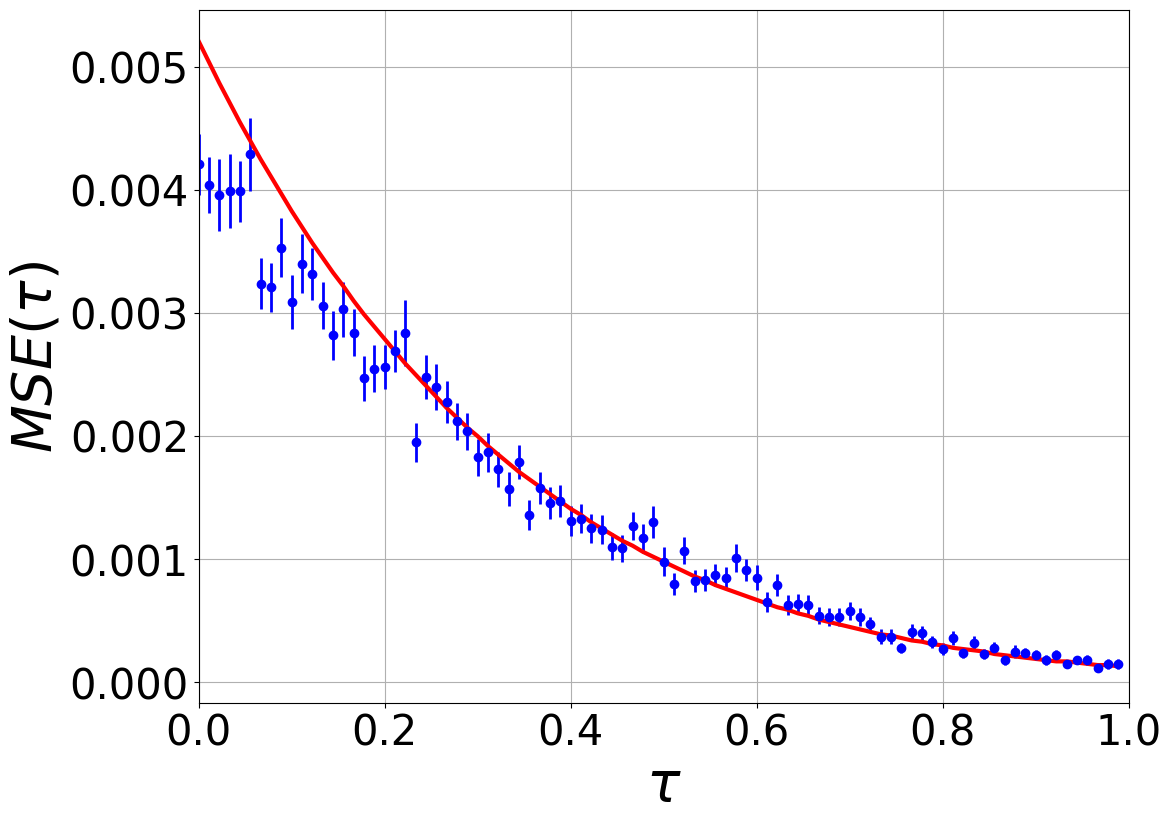}
    \caption{\small 
    Mean Square Error of decimation in the case of sparse Ising priors: theory versus Decimated AMP algorithm. The red solid curves are the expected pattern MSE predicted by theory as a function of the decimation time (i.e. the number of decoded patterns). The blue data points and error bars are obtained by running DAMP over $n=300$ independent instances. $N=1500$, $\lambda=0$ in all plots. \textbf{Left panel}: $\rho=1$, $\alpha=0.03$ namely $P=45$, $\Delta=0.08$ and $\beta=10$. \textbf{Middle panel}: $\rho=0.2$, $\alpha=0.04$ namely $P=60$, $\Delta=0.09$ and $\beta=8$. \textbf{Right panel}: $\rho=0.15$, $\alpha=0.06$ namely $P=90$, $\Delta=0.1$ and $\beta=8$.}
    \label{fig:MSE_vs_time}
\end{figure}

\begin{algorithm}[h]
\caption{Decimated AMP (DAMP)}\label{alg:decimated-AMP}
\footnotesize
\begin{algorithmic}
\Require N, P or $\alpha$, $\mathbf{Y}$, $\boldsymbol{\xi}$, $\epsilon$
\While{$\mu\leq P$}
    \State $\mathbf{g}\gets\mathcal{N}(0,\mathbbm{1}_N)$
    \State $\mathbf{x}^0\gets\sqrt{1-\epsilon^2}\mathbf{g}+\epsilon\boldsymbol{\xi}^\mu$
    \State $\mathbf{v}^0\gets 1-0.9(\mathbf{x}^0)^{\circ 2}$
    \State $\langle\boldsymbol{\eta}^\mu\rangle_{R=\mu-1},\langle(\boldsymbol{\eta}^\mu)^{\circ 2}\rangle_{R=\mu-1}-\langle\boldsymbol{\eta}^\mu\rangle_{R=\mu-1}^{\circ 2}\gets \text{AMP}(\mathbf{Y}_{R=\mu-1},\mathbf{x}^0,\mathbf{v}^0)$
    \State $\mathbf{Y}_{R=\mu}=\mathbf{Y}_{R=\mu-1}-\frac{\langle\boldsymbol{\eta}^\mu\rangle_{R=\mu-1}\langle\boldsymbol{\eta}^\mu\rangle_{R=\mu-1}^\intercal}{\sqrt{N}}$
\EndWhile
\State \textbf{Return}  $(\langle\boldsymbol{\eta}^\mu\rangle_{R=\mu-1},\langle(\boldsymbol{\eta}^\mu)^{\circ 2}\rangle_{R=\mu-1})_{1\leq \mu\leq P}$.
\end{algorithmic}
\end{algorithm}
It is a known fact, that in the Hopfield model AMP needs to be initialized sufficiently close to the patterns to converge, and here we experience the same behavior starting from the first step of decimation until the end. Hence DAMP is not suitable as an inference algorithm as it needs an informative initialization, whose correlation with the pattern sought is $\epsilon$ in Algorithm \ref{alg:decimated-AMP}. Nevertheless, DAMP can be considered as a tool to verify that our replica computations are correct and that decimation is able to retrieve all the patterns, which means it does not corrupt itself too much. 

In \figurename\,\ref{fig:MSE_vs_time} we plot the predicted theoretical curves of the expected MSE on the reconstruction on the single pattern
\begin{align}
    \EE\text{MSE}(\boldsymbol{\xi}^\mu;\boldsymbol{\eta}^\mu)=\frac{1}{N}\Vert\boldsymbol{\xi}^\mu-\langle\boldsymbol{\eta}^\mu\rangle_{t\mid tP=\mu-1}\Vert^2\simeq 1+q_t-2m_t
\end{align}
in red, where the subscript $t$ indicates that we at the decimation time $t$. The blue data points and error bars are obtained from an average of 300 instances of DAMP run on independently generated data. We considered different values of sparsity and the regularization parameter $\lambda$ was always set to $0$. In every case the theoretical curve seems to reproduce accurately the behaviour of the pattern MSE, yielding a good confirmation of our RS theory.


\subsection{Expected decimation performance}
In this section, we compare the expected denoising performance of decimation with the typical performance of a Rotation Invariant Estimator (RIE) introduced in \cite{RIE_bouchaud}. A RIE is characterized by the fact that it provides an estimate of the original matrix $\bxi \bxi^T$ which has the same eigenbasis as the one of the data matrix $\bY$. Once the eigenbasis is established, one only has to produce an estimate on the specturem based on that of $\bY$. As such, the RIE is a purely spectral estimator and it does not exploit the prior knowledge on the signal components. Among the possible RIEs, the one that acts optimally on the spectrum of $\bY$ is 
\begin{align}
    \hat{\boldsymbol{\lambda}}=\boldsymbol{\lambda}_{\bY}-2\Delta\mathcal{H}[\rho_\bY](\boldsymbol{\lambda}_{\bY})
\end{align}where $\hat{\boldsymbol{\lambda}}$ and $\boldsymbol{\lambda}_{\bY}$ are the vector of the eigenvalues of the estimate and of $\bY\sqrt{N}$ respectively, $\mathcal{H}[\rho_{\bY}]$ is the Hilbert transform of the spectral density of $\bY/\sqrt{N}$.

We shall measure the performance of an estimator $\bS$, whose eignevalues are of order $1$ by convention, with the matrix MSE:
\begin{align}
    \label{eq:matrix-MSE}
    \text{mMSE}(\bS;\bxi)=\frac{1}{N}\EE\Big\Vert\bS-\frac{\bxi\bxi^\intercal}{\sqrt{NP}}\Big\Vert^2_F\,,
\end{align}
and the matrix norm is the Frobenius' norm. The estimator produced by decimation would thus be
\begin{align}
    \label{DAMP-estimator}
    \bS_{\rm dec}:=\sum_{\mu=1}^P\frac{\langle\boldsymbol{\eta}^\mu\rangle_{R=\mu-1}\langle\boldsymbol{\eta}^\mu\rangle_{R=\mu-1}^\intercal}{\sqrt{NP}}
\end{align}
In order to make the comparison we need to connect the mMSE predicted by the theory for the decimation estimator with the definition \eqref{eq:matrix-MSE}, namely to re-express the latter in terms of the order parameters of the decimation free entropies. This can be done as follows, leveraging the assumption \eqref{local_measure}. By expanding the square in the mMSE definition evaluated at $\bS_{\rm dec}$ we recognize three main contributions:
\begin{align}
    &\frac{1}{N^2P}\sum_{i,j=1}^N\sum_{\mu,\nu=1}^P\EE[\xi^\mu_i\xi^\mu_j\xi^\nu_i\xi^\nu_j]=\frac{1+\alpha}{2}+o_N(1)\\
    \label{eq:2nd_mMSE_contribution}
    &\frac{1}{N^2P}\sum_{i,j=1}^N\sum_{\mu,\nu=1}^P\EE[\xi^\mu_i\langle\eta^\mu_j\rangle\xi^\nu_i\langle\eta^\nu_j\rangle]\\
    \label{eq:3rd_mMSE_contribution}
    &\frac{1}{N^2P}\sum_{i,j=1}^N\sum_{\mu,\nu=1}^P\EE[\langle\eta^\mu_i\rangle\langle\eta^\mu_j\rangle
    \langle\eta^\nu_i\rangle\langle\eta^\nu_j\rangle]
\end{align}where we dropped the subscrpts in the Gibbs brackets for convenience. While the first one can be computed right away using the properties of the prior, the other two require some extra effort. Concerning \eqref{eq:2nd_mMSE_contribution} we have:
\begin{align}
\begin{split}
    \frac{1}{N^2P}\sum_{i,j=1}^N\sum_{\mu,\nu=1}^P\EE[\xi^\mu_i\langle\eta^\mu_j\rangle\xi^\nu_i\langle\eta^\nu_j\rangle]&=
    \frac{1}{N^2P}\sum_{i,j=1}^N\sum_{\mu,\nu=1}^P\big[\delta_{\mu\nu}\xi^\mu_i\langle\eta^\mu_j\rangle\xi^\mu_i\langle\eta^\mu_j\rangle+\delta_{ij}\EE(\xi_i^\mu)^2\langle\eta^\nu_i\rangle^2\big]=\\
    &=\frac{1}{P}\sum_{\mu=1}^P(m^\mu)^2+\frac{\alpha}{P}\sum_{\mu=1}^P q^\mu+o_N(1)
\end{split}
\end{align}
where we have enforced \eqref{local_measure} and $q^\mu$ and $m^\mu$ are the overlap and Mattis magnetization respectively coming from the $\mu$-th decimation step. Let us now turn to \eqref{eq:3rd_mMSE_contribution}. Using similar arguments one can argue that:
\begin{align}
    \begin{split}
    \frac{1}{N^2P}\sum_{i,j=1}^N\sum_{\mu,\nu=1}^P\EE[\langle\eta^\mu_i\rangle\langle\eta^\mu_j\rangle
    \langle\eta^\nu_i\rangle\langle\eta^\nu_j\rangle]&=\frac{1}{P}\sum_{\mu=1}^P(q^\mu)^2+\alpha\Big(\frac{1}{P}\sum_{\mu=1}^P q^\mu\Big)^2+o_N(1)
    \end{split}
\end{align}
Therefore, collecting all the contributions one gets the asymptotic prediction:
\begin{align}
    \label{eq:RS-matrixMSE}
    {\rm mMSE}(\bS_{\rm dec};\bxi)\simeq \frac{1}{P}\sum_{\mu=1}^P\big(1+(q^\mu)^2-2(m^\mu)^2\big)+\alpha\Big(1-\frac{1}{P}\sum_{\mu=1}^P q^\mu\Big)^2\,.
\end{align}

\begin{figure}[t]
    \centering
    \includegraphics[width=0.325\textwidth]{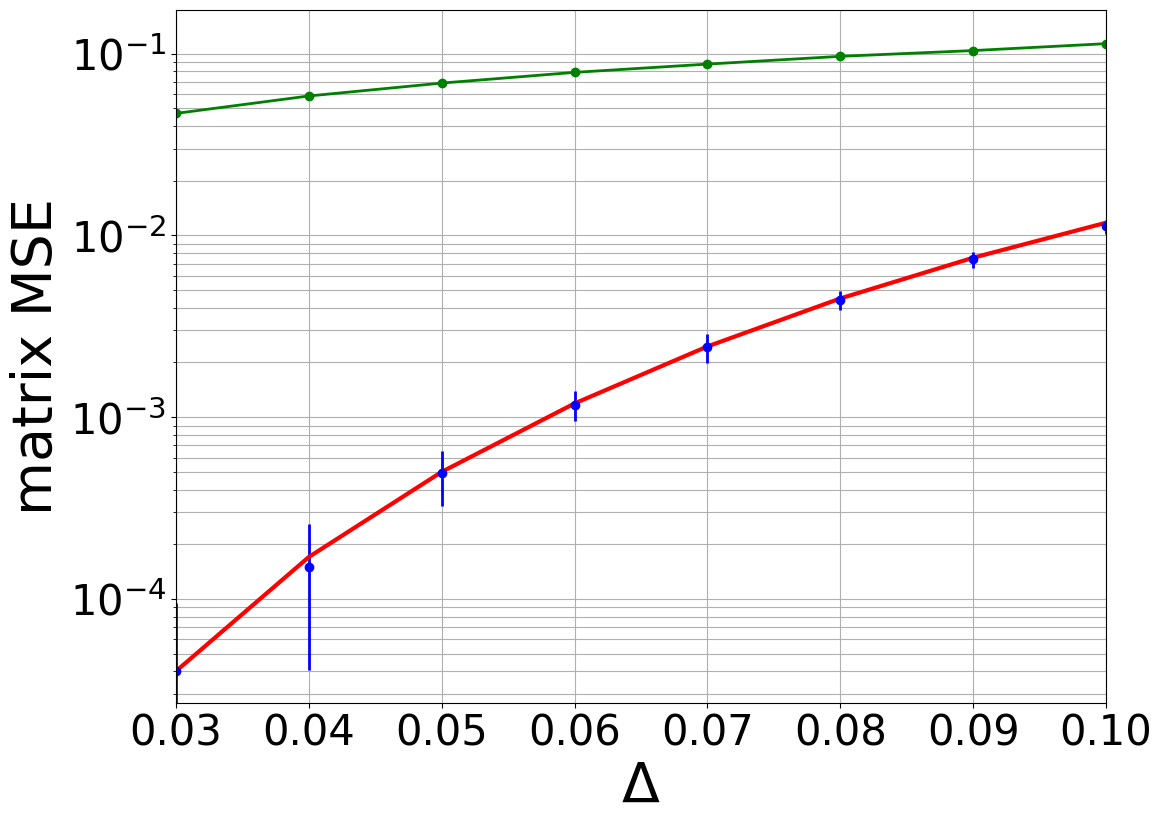}
    \includegraphics[width=0.325\textwidth]{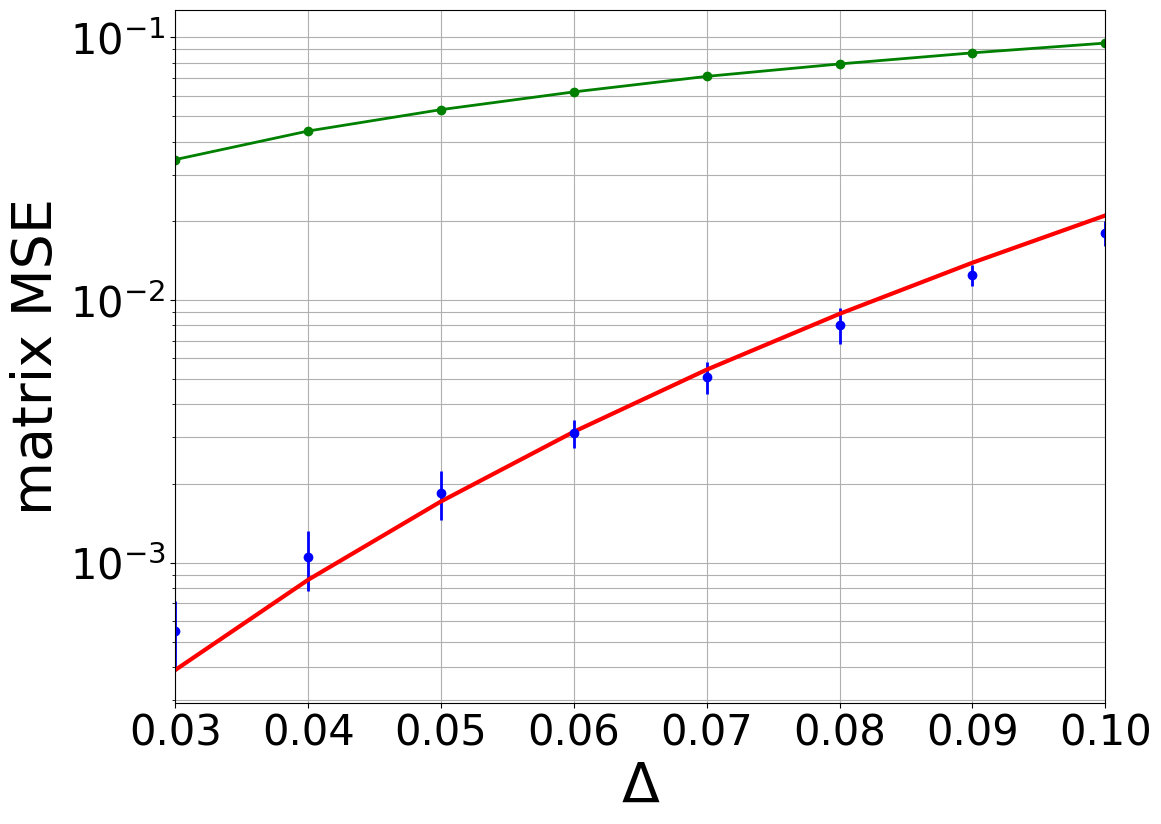}
    \includegraphics[width=0.325\textwidth]{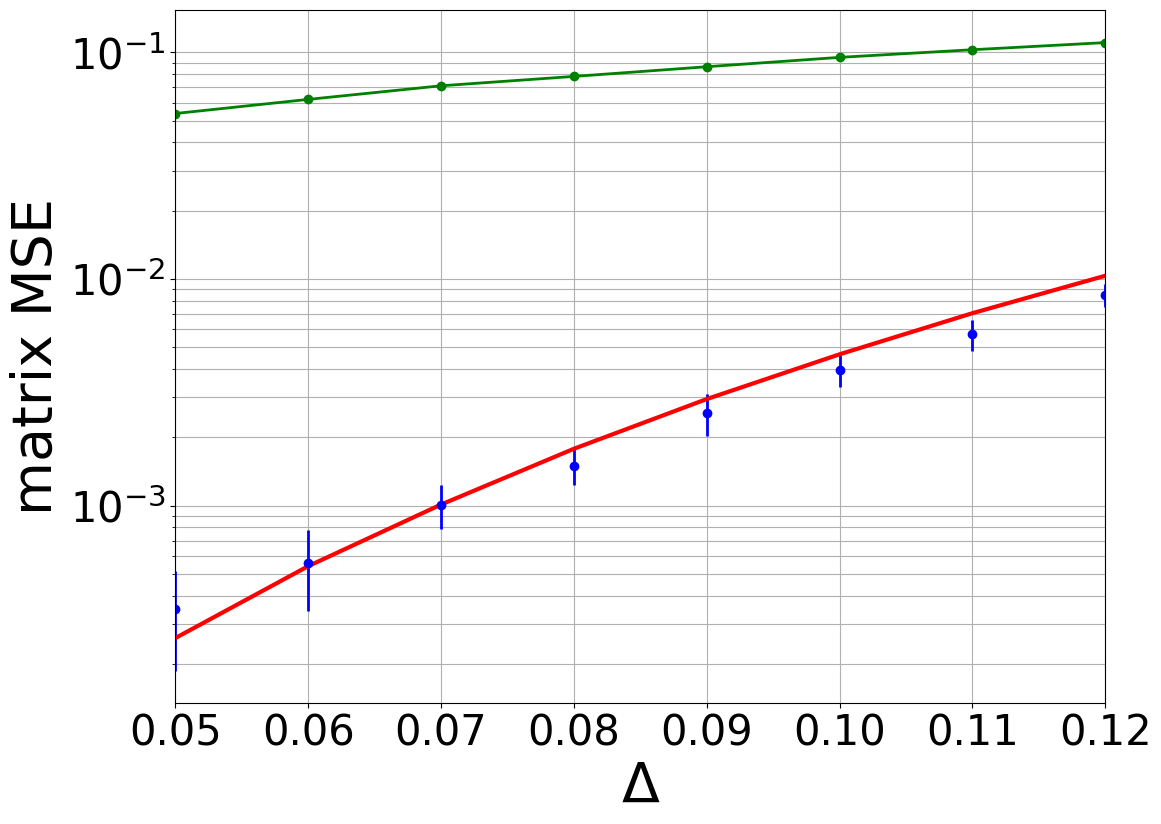}
    \caption{\small Matrix MSE as a function of $\Delta$ for sparse Ising priors with various sparsities. In green the denoising performance of a RIE, obtained by averaging over 30 independent samples. Error bars, corresponding to one standard deviation, are too small to be seen. In red, the performance predicted for an algorithm implementing decimation. The blue data points are obtained averaging over 30 DAMP's outputs, run on independently generated data. Error bars correspond to one standard deviation. In all cases $\lambda=0$, $\beta=8$ and $N=1500$. \textbf{Left panel}: $\rho=1$, $\alpha=0.03$ namely $P=45$ and $\Delta=0.08$. \textbf{Middle panel}: $\rho=0.2$, $\alpha=0.07$ namely $P=105$ and $\Delta=0.09$. \textbf{Left panel}: $\rho=0.15$, $\alpha=0.07$ namely $P=105$ and $\Delta=0.1$.   
    }
    \label{fig:matrixMSE_vs_RIE}
\end{figure}

In \figurename\,\ref{fig:matrixMSE_vs_RIE} we compare the performance of the RIE, in green, against the theoretical performance predicted for decimation in red, and the blue data points are obtained using the estimator produced by decimation (DAMP). As we can see there is a good agreement between DAMP and the theory, and both outperform the RIE as we expected. The RIE appears more robust to both noises (a) and (b), tuned by $\Delta$ and $\alpha$ respectively. On the contrary, the performance of decimation deteriorates quickly as soon as we get out of the retrieval region in the phase diagrams \figurename\,\ref{fig:phasediag_sparse}-\ref{fig:0T_phasediag_uniform}, and the amount of noise it can bear is strongly affected by the nature of the signal (sparse Ising or continuous). However, one must bear in mind that RIEs are suitable only for matrix denoising, and no information is reconstructed on the signal factor $\bxi$. Moreover, we notice that the performance of the RIE does not change sensibly from the left to the right panel ($\rho=1$ to $\rho=0.15$), and this is coherent with its purely spectral nature. In fact, the empirical spectral distribution of $\bxi\bxi^\intercal/\sqrt{NP}$ always converges to a Marchenko-Pastur law because of the completely factorized prior on the elements of $\bxi$. Hence, the small changes from the left to the right panel are mostly due to the slight increment in the noise level $\Delta$ and the aspect ratio (or load) $\alpha$.

\subsection{A ground state oracle for sparse Ising priors}
Our ground state oracle is based on an iterated simulated annealing (SA) routine that can be found in Algorithm \ref{alg:SA}, which is a refinement of the one in \cite{MFNN}.
\begin{algorithm}[h]
\caption{Simulated annealing (SA)}\label{alg:SA}
\footnotesize
\begin{algorithmic}
\Require $N$, $\bY$, threshold, $\beta_{\rm max}\in\mathbb{R}$, niter ($\in\mathbb{N}$), maxr ($\in\mathbb{N}$), restarts ($\in\mathbb{N}$)
\State itry $\gets 0$
\State found$\gets$\textbf{False}
\While{itry$ < 300$} \textbf{and} found$==$\textbf{False}
    \State stop$\gets 0$
    \State $\beta\gets 0$
    \State $\mathbf{s}\gets$ random sample from $\prod_{i=1}^NP_\xi$
    \State itry$\gets$itry$+1$
    \If{itry+restarts$>$maxr}
        \State \Return $\mathbf{s}$, itry
    \EndIf
    \If{itry$\%20=0$}
        \State threshold$\gets$threshold$\,\cdot\,0.9975$
    \EndIf
    
    \While{$k<$niter}
        \State $k\gets k+1$
        \State $\beta\gets1+\frac{k}{\rm niter}\,\cdot\,\beta_{\rm max}$
        \State $\mathbf{h}\gets \frac{\mathbf{Y}}{\sqrt{N}}\mathbf{s}$
        \State $V\gets \frac{\Vert\mathbf{s}\Vert^2}{N}+\frac{\lambda}{N}(\Vert\mathbf{s}\Vert^2-1)$
        \State $\mathbf{Z}_{\rm loc}\gets (1-\rho)\mathbf{1}+\rho\cosh(\beta\mathbf{h})e^{-\frac{\beta V}{2}}$\quad \quad \quad\quad\quad (Scalar functions are applied component-wise to vectors.)
        \State sample $\mathbf{ss}$ from ${\exp\big(\beta \mathbf{h}\cdot (\cdot)-\frac{\beta V}{2}\big)}/\mathbf{Z}_{\rm loc}$
        \If{$\Vert\mathbf{s}-\mathbf{ss}\Vert<10^{-3}$}
            \State $\mathbf{s}\gets\mathbf{ss}$
            \State stop$\gets$stop$+1$\quad\quad \quad\quad \quad\quad \quad\quad \quad \quad\quad\quad \quad (Updates become negligible.)
            \If{stop$>5$}
                \If{$-E(\mathbf{s}\mid\bY)>$threshold}
                    \State \Return $\mathbf{s}$, itry
                \Else 
                    \State\textbf{break}\quad \quad\quad \quad\quad \quad\quad \quad \quad\quad \quad\quad(wrong energy, try again)
                \EndIf
            \EndIf
        \Else
            \State stop$\gets 0$
            \State $\mathbf{s}\gets\mathbf{ss}$
        \EndIf
    
    \EndWhile
\EndWhile
\end{algorithmic}
\end{algorithm}

The energy landscape at the various steps of decimation is very similar to that of the Hopfield model. Consequently, algorithms that search for minima get frequently stuck in metastable states, which have a low overlap with the patterns. SA is not immune to this phenomenon. Therefore, we equip our SA routine with an acceptance criterion of the configuration output by the algorithm, that is based on the computation of the energy:
\begin{align}
    -E(\mathbf{s}\mid\mathbf{Y}_R)=\frac{1}{2\sqrt{N}}\mathbf{s}^\intercal\mathbf{Y}_R\mathbf{s}-\frac{\Vert\mathbf{s}\Vert^4}{4N}-\frac{\lambda}{4N}\big(\Vert\mathbf{s}\Vert^2-1\big)^2
\end{align}which is nothing the energy of our model at the $R$-th decimation step. Notice that this quantity is accessible by the Statistician and it is thus correct to use it as an input for a candidate algorithm. In Algorithm \ref{alg:SA} niter is the maximum number of temperature updates we allow, maxr is instead the maximum number of restarts allowed, considering also the restarts coming from previous pattern searches. The reason why we introduced this additional control is that typically when a bad configuration is accepted as a pattern estimate by mistake, the ensuing searches for other patterns require even more restarts. The above SA routine has to be combined with decimation, so once a configuration is accepted as a pattern the observations are modified $\bY\leftarrow \bY-\frac{\mathbf{s}\mathbf{s}^\intercal}{\sqrt{N}}$ and the routine is restarted. In order to make sure we really find patterns, we thus run all the algorithm (SA plus decimation) multiple times, typically five, and then we accept the output that required the least number of restarts to be produced. This procedure is costly, and as noticed already in \cite{MFNN}, it requires an exponential number of restarts.

Algorithm \ref{alg:SA} suffers from the same issues as the one in \cite{MFNN}. For instance, the overall decimation procedure still requires an exponential (in $N$) number of restarts. However, the presence of sparsity introduces further non-trivial complications. In fact, the signal components are no longer constrained on the hypercube, and this allows for fluctuations in the norm of the outputs that reflect in fluctuations on the average energy of the patterns. Specifically, the more sparse the signal is, the wider the gap between the highest and the lowest energy of the patterns. These fluctuations can challenge the energy restarting criterion in our SA routine, that can thus confuse a metastable state for a pattern.

\begin{figure}[t]
    \centering
    \includegraphics[width=0.325\textwidth]{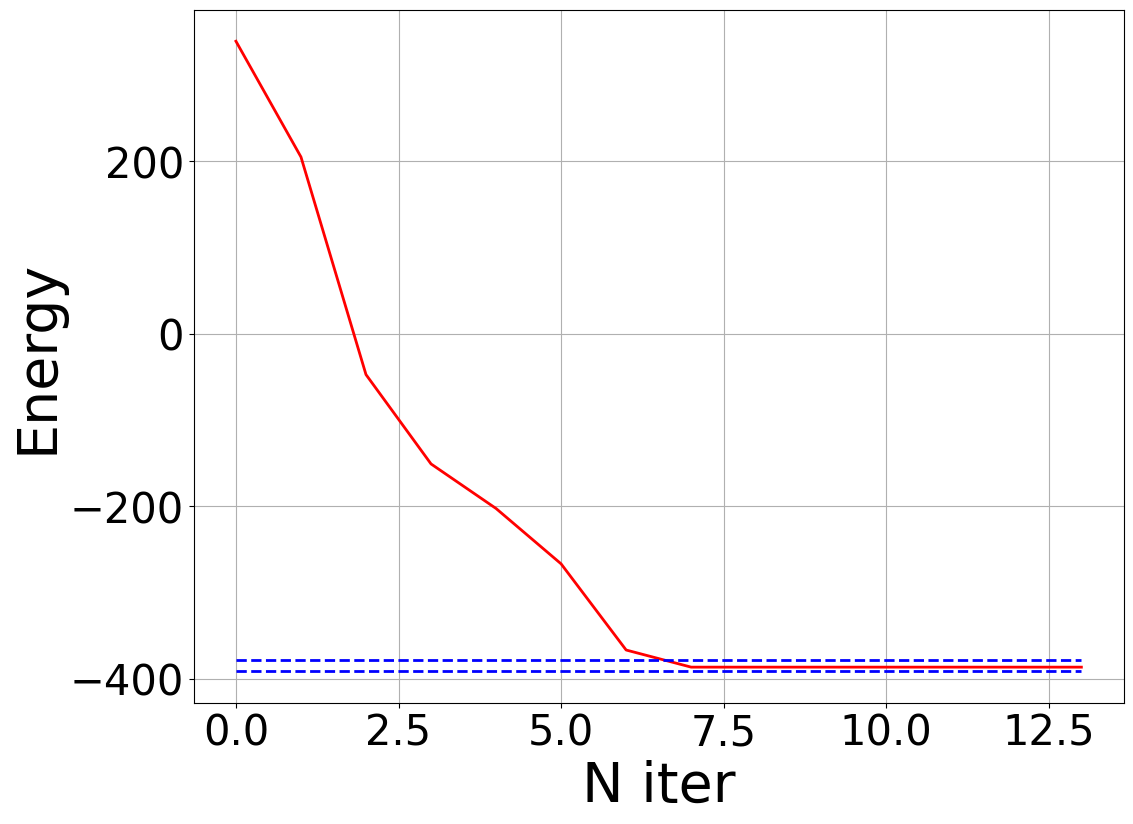}
    \includegraphics[width=0.325\textwidth]{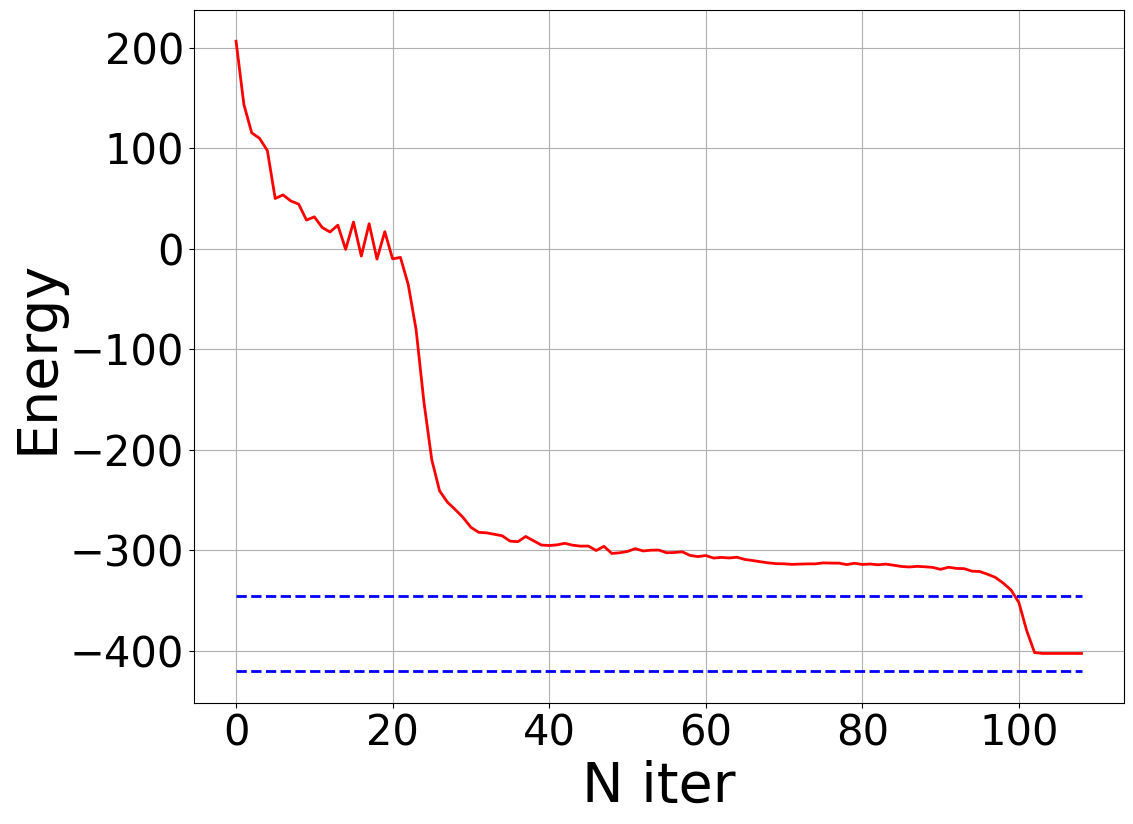}
    \includegraphics[width=0.325\textwidth]{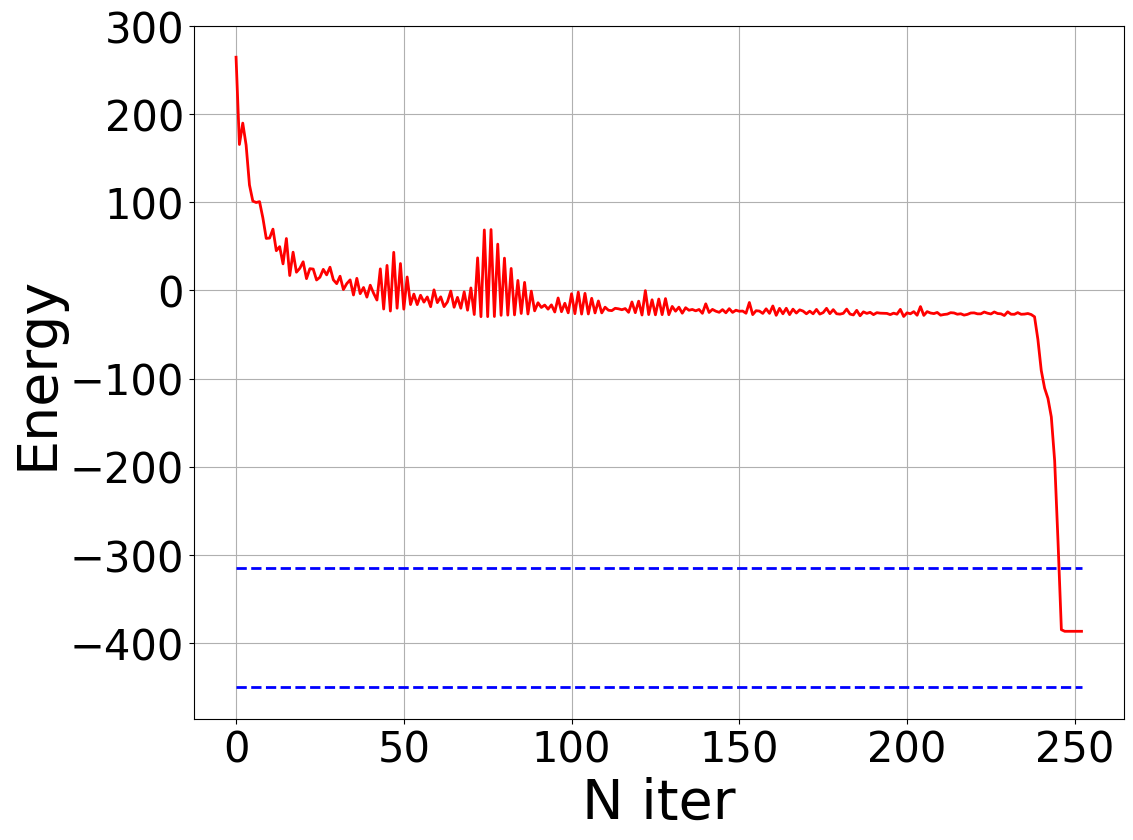}
    \caption{\small Energy landscape exploration of the Simulated Annealing applied to sparse Ising priors. On the vertical axis we have the energy value as a function of the number of iterations (temperature updates) of SA on the horizontal axis. For all the three plots $N=1500$, $\alpha=0.01$ (namely only 15 patterns to be found), $\Delta=0.05$ and $\lambda=-0.08$. From the left to the right: $\rho=1,0.3,0.15$. The patterns were reconstructed exactly in all thre cases. SA finds immediately the patterns for low sparsities $\rho\sim1$. As soon as sparsity increases, a lot of configurations start to exhibit an almost vanishing energy (recall that the noise shifts this value). The dashed blue lines mark the highest and the lowest pattern energy. As we can see the band they identify is narrow with low sparsity, and it becomes wider for higher values of sparsity due to more intense fluctuations.}
    \label{fig:energy_landscapes}
\end{figure}

Furthermore, one observes that when too few patterns are stored or remain in $\bY$, it is harder for the SA routing to find them. If, for instance, we only have one pattern left, the Hebbian matrix $\bxi\bxi^\intercal$, which is supposed to attract the $\bx$-configurations towards the pattern, has only a fraction $\rho^2$ of non-zero components. This gives rise to a large number of configurations that have degenerate energy, close to $0$. The energy landscape thus appears as a golf course, flat almost everywhere, except for a pit, corresponding to the pattern left. From our numerical experiments, this effect seems to hold also for more than one, but still few, patterns stored. See \figurename\,\ref{fig:energy_landscapes}.

\subsection{Reversed decimation}
In all the tests we have run, the performance of decimation in reconstructing the patterns improves along the procedure itself. The last patterns are always better estimated than the first ones, and this supports the idea that decimation effectively decreases the pattern interference. In particular, it is clear that the quality of reconstruction of one pattern depends on the previous \enquote{history} of the process.

\begin{figure}[h]
    \centering
    \includegraphics[width=0.6\textwidth]{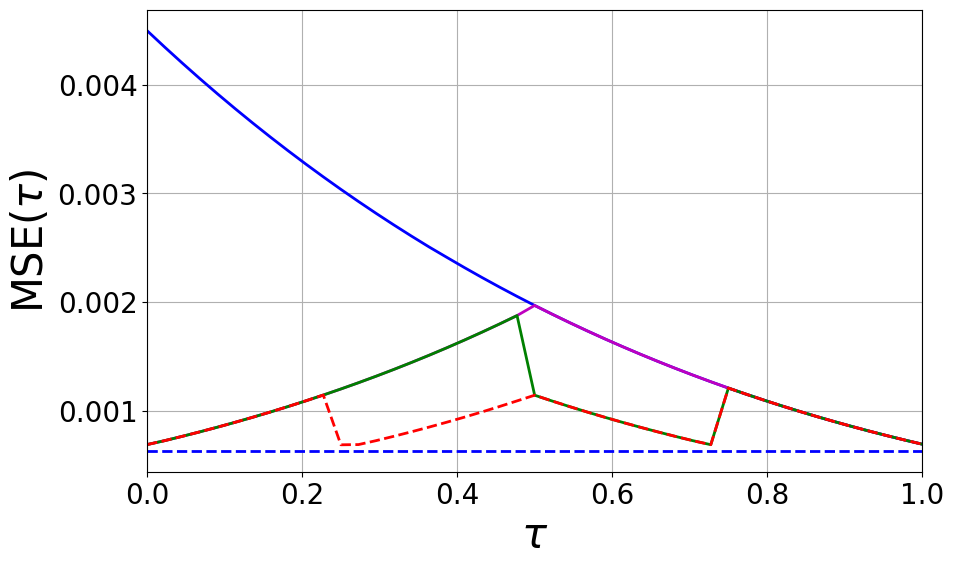}
    \caption{\small Improvement in performance obtained re-iterating decimation for Rademacher prior. In this example $\Delta=0.08$, $\alpha=0.03$, $\rho=1$ and $\beta=10$. The blue line is the first run, where the expected MSE on the reconstruction of the single patterns decreases along decimation. The magenta curve is instead obtained by fixing the last half of pattern MSEs, and running decimation backwards. Starting from the magenta line, we obtained the green solid line by fixing the first half and the last quarter of MSEs, and then running decimation for finding the third
    quarter of MSEs. Finally, the red dashed line was obtained from the green line running decimation again, with fixed first quarter and last half of MSEs. The blue dashed line is the expected MSE predicted by the rank one formula. Coherently, the last decimation steps approach the rank-one formula MSE from above, because the interference noise has been almost completely eliminated, except for noise of decimation itself, that is responsible for the final small gap.}
    \label{fig:rev-decimation}
\end{figure}

Once the procedure exhausts the patterns, one can imagine to run it again backwards, keeping the last half of the patterns that were reconstructed with higher accuracy. As illustrated in \figurename\,\ref{fig:rev-decimation}, this improves the reconstruction performance also for the first half of the patterns. One can then re-iterate the same procedure, keeping only the first $1/2$ and the last $1/4$ of the patterns, that are now the best reconstructed ones. This in turn leads to a further improvement in the reconstruction also for the middle patterns. This reasoning can be iterated ad libitum. 

In \figurename\,\ref{fig:rev-decimation} we see how performance improves in the various rounds of decimation, and we compare it to the performance predicted by the rank-one formula, i.e. what we should have for any sub-linear rank ($\alpha=0$, see Section \ref{sec:related_works}). We see that, little by little, the performance approaches that of the rank-one formula.

\section{Related works}\label{sec:related_works}

\subsection{Unlearning and dreaming} As evident from \figurename\,\ref{fig:phasediag_sparse}, without having strong sparsity, the storage capacity of the model is not very large, and the network is far from being able to store an over-complete basis of $\R^N$. In an attempt to solve this issue one can pre-process the observation matrix with Hebbian unlearning \cite{Hopfield_unlearning,vanHemmen_unlearning}, with which decimation itself bears some similarity. Unlearning consists in iterating a zero temperature dynamics until convergence, which is likely to occur at a spurious state $\boldsymbol{\eta}$ that is then removed from the observations $\bY\leftarrow\bY-\varepsilon\boldsymbol{\eta}\boldsymbol{\eta}^{\intercal}/\sqrt{N}$, with a small $\varepsilon$. If run for an appropriate number of times, unlearning acts on the energy landscape penalizing spurious metastable states. This procedure has two fundamental parameters to be tuned: $\varepsilon$ and the number of times $D$ it is iterated \cite{Benedetti_unlearning}. If $\varepsilon$ or $D$ are too large one risks to remove also the wanted patterns.

Apart from numerical evidence, there is little theoretical understanding of the unlearning procedure as illustrated above. However, there are other convenient iterative ways of modifying the Hebbian matrix \cite{Dotsenko_1991,Plakhov_unlearning,Agliari-alemanno_dreaming_NNs,alemanno_dreaming_RBM} that converge to the so called pseudo-inverse learning rule (or modifications of it) \cite{Kohonen1988SelfOrganizationAA,personnaz,Kanter_sompo_pseudoinv}, which in turn is able to increase the storage capacity to $\alpha_c=1$.

Despite the apparent similarities, the goal of decimation is very different from that of unlearning. Its aim is to find a pattern, and not a metastable state, and to remove it completely (or almost completely) from $\bY$, which amounts to set $\varepsilon=1$ (or close to $1$) above. Furthermore, it is worth stressing that, unlike classical unlearning, we have a theoretical control on decimation, namely we can track its behaviour step by step.

\subsection{Sub-linear rank} In a recent work \cite{Jean_sublinear} the authors discuss the denoising of large matrices in the same setting as ours, with a main focus on the case $P=N^\delta$, $\delta\in(0,1)$, i.e. a sub-linear rank regime. In the mentioned paper, it is stated that, as long as the prior on the $N\times P$ matrix $\bxi$ is completely factorized over the matrix elements, the mutual information between $\bxi$ and the data is given by the rank-one replica formula for \emph{any} sub-linear rank regime, in agreement with \cite{justin_Husson_sublinear_sphint}. Though not explicitly stated in our previous work \cite{MFNN}, our findings indeed suggest the same result, as it can be deduced from Section \ref{sanity_checks_section}. In fact our free entropy, which is in close relation with the mutual information between observations and signal, takes the same form for any $P$ such that $P/N\to0$. Furthermore, for $\alpha=0$ and $\beta=1/\Delta$, the fixed point equations admit a self-consistent solution that satisfies the Nishimori identities, which suggests that Bayes-optimality is recovered. From the form of the free entropy \eqref{RS_free_entropy}, it is also evident that the effect of decimation is visible only for truly extensive rank. The reason is that, if we penalize a finite number of directions in a space of dimension growing to infinity, the system can easily find other favoured directions to termalize in. In other words, the $p^\mu(\mathbf{x})$'s in \eqref{eq:generic_step_energy} give a sub-extensive contribution that can be neglected in any sub-linear rank regime.

Another delicate point is the definition of DAMP. We stress that in \eqref{DAMP:cavityfieldA} and \eqref{DAMP:cavityfieldB} the presence of a high-rank spike inside $\bY$ can induce non-trivial modifications both in $\bA$ and $\bB$. More specifically, it is known that, for instance, the Onsager reaction in \eqref{DAMP:cavityfieldA} containing $\bY^{\circ 2}$ has different asymptotically equivalent formulations. In the case of a Gaussian channel with a low-rank spike $\bY^{\circ 2}$ can be replaced by an all-ones matrix. This is due to the fact that the rank of the spike is not large enough to induce modifications in the spectrum of the noise matrix. In the high-rank regime, on the contrary, the extensive rank starts to play a role and gives rise to important contributions in the reaction term. Moreover, the reaction term changes also along the decimation procedure, in which one further perturbs the data matrix with the high rank matrix of the decimation estimates $\sum_{\mu=P-R+1}^P\frac{\boldsymbol{\eta}^\mu\boldsymbol{\eta}^{\mu\intercal}}{\sqrt{N}}$. Hence, the formulation in \eqref{DAMP:cavityfieldA}-\eqref{DAMP:cavityfieldB} turns out to be convenient. The low-rank regime is insensitive to the aforementioned changes.

Despite we were not able to prove it, \figurename\,\ref{fig:rev-decimation} suggests that re-iterating decimation in a  proper way could lead to a performance similar to that predicted by the low rank replica symmetric formula. One may be led to think that reversed decimation yields Bayes-optimal performance. This is however not true. In fact, in the high rank case the spike induces a non-negligible perturbation of the spectrum of the noise matrix that can be used to perform inference (this deformation is captured by the RIE for instance) especially for large $\alpha$'s, where decimation fails.

\subsection{Channel universality properties} Low-rank spiked models are known to fulfill channel universality \cite{channel_universality,guionnet2022low,Justin_Alice_mismatch}, namely for any well-behaved $P_{\rm out}(y\mid x)$ and data generated with the rule
\begin{align}
    \label{P_out_spiked}
    Y_{ij}\sim P_{\rm out}\Big(\cdot\mid\sum_{\mu=1}^P\frac{\xi^\mu_i\xi^\mu_j}{\sqrt{N}}\Big)
\end{align}
the mutual information between the data $\bY$ and $\bxi$ can be computed through an equivalent Gaussian channel as in \eqref{eq:channel} with a properly tuned noise intensity $\Delta$. The proof of this equivalence requires two concomitant behaviours, \emph{i)} universality in the likelihood, and \emph{ii)} universality in the quenched disorder (i.e. the law of the data $\bY$), and holds as long as $P^3/\sqrt{N}\to0$ \cite{guionnet2022low}. Informally, the main idea is to expand $P_{\rm out}\Big(\cdot\mid\sum_{\mu=1}^P\frac{\xi^\mu_i\xi^\mu_j}{\sqrt{N}}\Big)$ around $0$ in its second entry up to second order, since for low-rank spikes $\sum_{\mu=1}^P\frac{\xi^\mu_i\xi^\mu_j}{\sqrt{N}}$ is small for any fixed couple of indices $i,j$. On the contrary, in the high-rank setting the higher moments of the spike start to matter, meaning that the previous expansion fails, and universality breaks down.

In our mismatched setting one can still count on the universality of the likelihood \emph{for a single decimation step}. In fact, here the Statistician assumes to observe a low-rank spike, that is they consider
\begin{align}
    Y_{ij}\sim P_{\rm out}\Big(\cdot\mid\frac{x_ix_j}{\sqrt{N}}\Big)
\end{align}
whereas the data are generated through \eqref{eq:channel}. The free entropy of the related model reads as
\begin{align}
    \frac{1}{N}\EE[\log\mathcal{Z}_R-\sum_{i,j}\log P_{\rm out}(Y_{ij}\mid 0)]=\frac{1}{N}\EE\log\int dP_\xi(\bx)\exp\Big[\sum_{i,j}\Big(\log P_{\rm out}\Big(Y_{ij}\mid 
    \frac{x_ix_j}{\sqrt{N}}
    \Big)-\log P_{\rm out}(Y_{ij}\mid 0)\Big)\Big]
\end{align}
where $\sum_{i,j}\log P_{\rm out}(Y_{ij}\mid 0)$ has been subtracted to have a proper scaling. From the above equation one readily realizes that an expansion up to second order of $P_{\rm out}$ yields the desired equivalent quadratic model, for which our computations hold. However, we stress that exploiting this universality produces errors of $O(N^{-1/2})$. These errors accumulate along the $P=O(N)$ steps of decimation resulting in potentially non-negligible deviations from the original model towards the end of the procedure.

\section{Conclusion and outlooks}\label{sec:conslusions}
Building on the results of \cite{MFNN}, we have extended the analysis of the decimation procedure to a wide class of priors on the matrix elements of the factors $\bxi$ for symmetric matrix factorization. We provided exhaustive numerical evidence in support of our replica theory, via the introduction of DAMP, whose performance in pattern retrieval, and matrix denoising matches the one predicted by the theory. Our numerical experiments confirm that decimation is a viable strategy for matrix factorization. In particular, as long as the first step is feasible, i.e. the procedure is started at a point of the phase diagram where there is a non-vanishing Mattis magnetization with one of the patterns, decimation is able to find all of them, up to a permutation. We stress again that DAMP is not an appropriate algorithm for inference, since it needs a strongly informative initialization. Nevertheless, in the case of sparse Ising priors,
we were able to find a ground state oracle that is able to find all the patterns in suitable regions of the phase space of the decimation neural network models. The latter still suffers from an exponential complexity: it needs an exponential number of restarts (in $N$) in order to find all the patterns and discard correctly the spurious states it may get stuck in.

The idea of reversed decimation and unlearning are insightful perspectives. In fact, in order to increase the storage capacity of the neural networks, or equivalently to widen the region of the phase space where we can perform matrix factorization, one could pre-process the Hebbian interaction matrix using a local updating rule, as the ones described in \cite{Agliari-alemanno_dreaming_NNs,Dreaming_NNs0}. In these works, besides the usual \enquote{forgetting} mechanism, the authors also consider a consolidation of the memories, which avoids the risk of corrupting the Hebbian interaction too much. This pre-processing could be combined with reversed decimation in order to obtain a better performing procedure that is also more robust to pattern interference.

Finally, in an upcoming work, we shall tackle the asymmetric problem, which is closer to practical applications. Here, the Statistician has to reconstruct two independent matrices $\bF\in\R^{N\times P}$ and $\bX\in\R^{P\times M}$ from the observations
\begin{align}
    \bY=\frac{1}{\sqrt{N}}\bF\bX+\sqrt{\Delta}\bZ\in\R^{N\times M}
\end{align}
in the scaling limit $N,M,P\to\infty$ with $P/N=\alpha>0$ and $P/M=\gamma>0$.

\small
\section*{Acknowledgments}
We would like to thank Enzo Marinari and Federico Ricci-Tersenghi for their suggestions on the reversed decimation, Enzo Marinari and Marco Benedetti for discussions on unlearning, as well as Florent Krzakala, Lenka Zdeborov\'a and Jean Barbier for many fruitful discussionson matrix factorization.  MM acknowledges financial support by the PNRR-PE-AI FAIR project funded by the NextGeneration EU program.

\printbibliography

\end{document}